\documentclass[%
reprint,
superscriptaddress,
amsmath,amssymb,
aps,
pra,
floatfix,
]{revtex4-2}

\usepackage{graphicx}
\usepackage{dcolumn}
\usepackage{bm}
\usepackage[hidelinks]{hyperref}

\usepackage{xcolor}
\hypersetup{
	colorlinks   = true, 
	urlcolor     = blue, 
	linkcolor    = blue, 
	citecolor   = blue 
}
\usepackage[T1]{fontenc} 
\usepackage{siunitx} 
\usepackage{soul} 

\usepackage{mathptmx} 
\usepackage[utf8]{inputenc} 

\providecommand{\abs}[1]{\lvert#1\rvert} 

\graphicspath{ 
	{figures} 
}
\makeatletter
\def\input@path{
	{figures/} 
} 
\makeatother

\newcommand{\Qf}{Q}
\newcommand{\Gp}{\Gamma_P}
\newcommand{\Gpuf}{\Gamma_{P\text{, unfiltered}}}
\newcommand{\Tp}{T_{1,P}}
\newcommand{\Tpt}{\tilde{T}_{1,P}}
\newcommand{\ropt}{r_\text{opt}}
\newcommand{\Pin}{P}
\newcommand{\wqt}{\omega_{q\text{,target}}}
\newcommand{\ah}{\hat{a}}
\newcommand{\gam}{\gamma}
\newcommand{\ain}{\hat{a}_\text{in}}
\newcommand{\bin}{\hat{b}_\text{in}}
\newcommand{\ndot}{\dot{n}}

\begin{document}
	
	\preprint{APS/123-QED}
	
	\title{Interferometric Purcell suppression of spontaneous emission in a superconducting qubit}
	
	\newcommand{\RLE}{\affiliation{Research Laboratory of Electronics, Massachusetts Institute of Technology, \\ Cambridge, Massachusetts 02139, USA}}
	\newcommand{\EECS}{\affiliation{Department of Electrical Engineering and Computer Science, Massachusetts Institute of Technology, \\ Cambridge, Massachusetts 02139, USA}}
	\newcommand{\LL}{\affiliation{MIT Lincoln Laboratory, Lexington, Massachusetts 02421, USA}}
	\newcommand{\Harvard}{\affiliation{Harvard John A. Paulson School of Engineering and Applied Sciences, Harvard University, \\ Cambridge, Massachusetts 02138, USA}}
	\author{Alec~Yen}\email{alecyen@mit.edu}\EECS\RLE
	\author{Yufeng~Ye}\EECS\RLE
	\author{Kaidong~Peng}\EECS\RLE
	\author{Jennifer~Wang}\EECS\RLE
	\author{Gregory~Cunningham}\RLE\Harvard
	\author{Michael~Gingras}\LL
	\author{Bethany~M.~Niedzielski}\LL
	\author{Hannah~Stickler}\LL
	\author{Kyle~Serniak}\RLE\LL
	\author{Mollie~E.~Schwartz}\LL
	\author{Kevin~P.~O'Brien}\email{kpobrien@mit.edu}\EECS\RLE
	
	\date{\today}
	
	\begin{abstract} 
		In superconducting qubits, suppression of spontaneous emission is essential to achieve fast dispersive measurement and reset without sacrificing qubit lifetime.
		We show that resonator-mediated decay of the qubit mode to the feedline can be suppressed using destructive interference, where the readout resonator is coupled to the feedline at two points. 
		This ``interferometric Purcell filter'' does not require dedicated filter components or impedance mismatch in the feedline, making it suitable for applications such as all-pass readout. 
		We design and fabricate a device with the proposed scheme and demonstrate suppression of resonator-mediated decay that exceeds 2 orders of magnitude over a bandwidth of $\SI{400}{MHz}$ for a resonator linewidth of \SI{13.8}{MHz}.
	\end{abstract}
	
	\maketitle
	
	\section{Introduction}
	Superconducting qubits have become a strong candidate platform for quantum computing, demonstrating a high degree of engineerability. Quantum non-demolition measurements can be performed by coupling the qubit dispersively to a readout resonator \cite{blais_cavity_2004,wallraff_strong_2004,wallraff_approaching_2005,gambetta_qubit-photon_2006}.
	The qubit imparts a state-dependent frequency shift to the resonator, which can then be measured by a microwave readout tone to determine the qubit state.
	Fast measurement and reset are critical for applications such as quantum error correction to minimize qubit decoherence and increase the code's repetition rate \cite{acharya_suppressing_2023, krinner_realizing_2022, chen_exponential_2021,bengtsson_model-based_2024}.
	This need has motivated the design of readout resonators with large linewidths, or decay rates, so that qubit state information is extracted as quickly as possible during readout.
	However, this also increases the rate at which the qubit loses information to the environment in the form of Purcell decay \cite{purcell_proceedings_1946}.
	To avoid this trade-off, Purcell filters have been introduced that engineer the admittance seen by the qubit through the readout resonator to be high at the resonator frequency but low at the qubit frequency \cite{reed_fast_2010, jeffrey_fast_2014,sete_quantum_2015}.
	
	Many methods of Purcell protection have been proposed for superconducting qubits. 
	A common approach inserts a low-quality factor (low-Q) bandpass filter between the readout resonator and the feedline \cite{jeffrey_fast_2014,sete_quantum_2015}.
	The use of dedicated filter modes for Purcell suppression is widespread; there have been many demonstrations of bandstop \cite{reed_fast_2010} and multi-stage bandpass filters \cite{bronn_broadband_2015,cleland_mechanical_2019,yan_broadband_2023,park_characterization_2024}.
	Other works have shown that Purcell decay can be suppressed with the addition of a second resonator mode \cite{govia_enhanced_2017} or a precise sub-femtofarad capacitance \cite{bronn_reducing_2015}.
	Recently, an ``intrinsic Purcell filter'' has been proposed \cite{sunada_fast_2022}, which carefully positions the readout resonator such that the coupling strength between the dressed-qubit mode in the readout resonator and the feedline is minimized. 
	Standing waves in the feedline are necessary for this scheme; in \cite{sunada_fast_2022}, the feedline is terminated in an open, so that the resonator-feedline coupling is purely capacitive and couples to the node of the dressed-qubit mode's electric field.
	
	Ultimately, the chosen form of Purcell suppression depends on the readout scheme.
	In a previous work, we proposed and demonstrated a high-fidelity ``all-pass readout'' scheme, using an all-pass resonator that preferentially emits readout photons toward the output \cite{yen_directional_2024}.
	This transmission-based scheme preserves preferential directional decay of the readout signal without using intentional mismatch and aims to reduce both (1) variation in resonator linewidth and (2) infrastructure overhead associated with impedance matching.
	We refer the reader to \cite{yen_directional_2024} for an in-depth quantitative discussion of the disadvantages of impedance mismatch in conventional schemes and how these are addressed by all-pass readout.
	However, because all-pass readout relies on the degeneracy of two resonator modes, it is not feasible to introduce additional low-Q bandpass filter or resonator modes, since this breaks the degeneracy condition.
	In addition, since all-pass readout is motivated by the removal of impedance mismatch in the feedline, it is also incompatible with the intrinsic Purcell filter, whose operation relies on the standing waves created by such mismatch \cite{sunada_fast_2022}.
	The proposal in \cite{bronn_reducing_2015} to cancel Purcell decay by adding a capacitor is challenging to implement since it requires a precise sub-femtofarad capacitance.
	Alternatively, one could place dedicated filter components at the input and output of the feedline, but this significantly increases both the footprint size and design complexity.
	Instead, we seek a form of Purcell suppression that both avoids adding filter components and is compatible with our implementation of all-pass readout.

	Here, we propose and demonstrate interferometric suppression of Purcell decay in a superconducting qubit.
	We couple the quarter-wavelength readout resonator to the feedline at two spatially separate points, such that the qubit mode destructively interferes in the feedline.
	In a planar circuit platform, we demonstrate suppression of resonator-mediated qubit decay by 2 orders of magnitude over a bandwidth of $\SI{400}{MHz}$, where Purcell suppression is defined as the ratio of the Purcell decay without the filter to the Purcell decay with the filter. 
	Biasing the flux-tunable transmon qubit to a notch frequency, we extract a Purcell-limited lifetime of more than \SI{16}{ms}, corresponding to a Purcell suppression factor of over two thousand.
	Our demonstrated filter enables applications such as all-pass readout with Purcell protection for robust and scalable quantum measurement.
	
	\section{Device Design}
	\begin{figure}[t]
		\centering
		\includegraphics{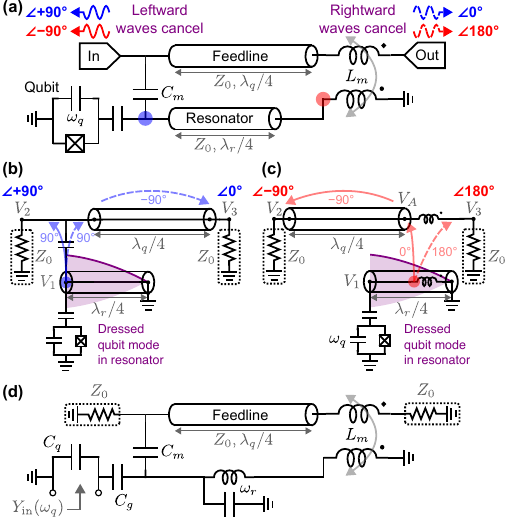}
		\caption{Proposed interferometric Purcell filter. (a) Circuit model, assuming $\wqt<\omega_r$. Lumped circuit approximations are used to represent the capacitive and inductive couplings.
			(b-c)~The circuit is decomposed into sub-circuits of either only (b) mutual capacitance or (c) mutual inductance.
			The phase of $V_1$ represents the phase of the dressed-qubit mode in the resonator.
			The labeled $\angle\theta$ associated with $V_2$ and $V_3$ represents the temporal phase $\theta$ relative to $V_1$. 
			Annotations in blue and red indicate voltage contributions from the capacitive and inductive coupling points, respectively.
			Solid and dashed arrows indicate voltage waves propagating to the left and the right, respectively.
			Superimposing the sub-circuits, we see there is destructive interference of the qubit mode at both ends of the feedline.
			(d) The circuit used for calculation of the Purcell decay rate using the admittance-based approach.}
		\label{fig:1}
	\end{figure}
	
	The unfiltered Purcell decay rate of a qubit capacitively coupled to a detuned lossy, single-mode readout resonator can be approximated as (see Appendix~\ref{sec:app-unfiltered})
	\begin{equation}
		\begin{aligned}
			\Gpuf&=\kappa\left(\frac{g}{\Delta}\right)^2\left(\frac{\omega_q}{\omega_r}\right)^3\left(\frac{2\omega_q}{\omega_q+\omega_r}\right)^2\\
			&\approx\kappa\left(\frac{g}{\Delta}\right)^2\left(\frac{\omega_q}{\omega_r}\right)^4,
		\end{aligned}
		\label{eq:Gp_unfiltered}
	\end{equation}
	where $\kappa$ is the resonator linewidth, $g$ is the frequency-dependent resonator-qubit coupling, $\Delta=\omega_q-\omega_r$ is the detuning between the qubit and resonator.
	We note that this expression includes a factor of $\sim\!\!(\omega_q/\omega_r)^4$ that is absent in the standard Purcell decay approximation given by $\Gp=\kappa(g/\Delta)^2$ \cite{blais_cavity_2004,koch_charge-insensitive_2007}.
	For a resonator frequency of $\omega_r=\SI{7}{GHz}$ and typical negative detunings of $\Delta=\SI{-1.5}{GHz}$ and $\Delta=\SI{-2.5}{GHz}$, we find that the standard expression overestimates the unfiltered Purcell decay by a factor of $2.7\times$ and $6.1\times$, respectively.
	We thus emphasize the importance of using the more realistic expression in Eq.~\eqref{eq:Gp_unfiltered} to provide a fairer estimate of the Purcell suppression achieved by a given filter.
	See Appendix~\ref{sec:app-unfiltered} for the analytic derivation of Eq.~\eqref{eq:Gp_unfiltered} and its agreement with circuit simulation.
	In all expressions throughout, the resonator-qubit coupling strength $g$ is assumed to have frequency dependence $g=g_0\sqrt{\omega_q/\omega_r}$, where $g_0$ is the coupling strength when the qubit and resonator are on resonance.
	
	The circuit model of the interferometric Purcell filter is depicted in Fig.~\ref{fig:1}a, consisting of a feedline, a quarter-wavelength transmission line resonator, and a transmon qubit.
	The feedline is coupled to both the open and shorted ends of the resonator, where the two coupling points are separated by a quarter-wavelength at the target qubit frequency $\wqt$.
	The open (shorted) end of the resonator couples capacitively (inductively) to the feedline as the antinode of the voltage (current) standing wave.
	In Fig.~\ref{fig:1}, we approximate these couplings as lumped mutual inductance (with coupling coefficient $k=1$) and mutual capacitance.
	A transmon qubit is coupled capacitively to the open end of the resonator.
	Using the same terminology as \cite{sunada_fast_2022}, we define a ``dressed-qubit mode'' within the resonator, where the transmon qubit is approximated as a linear resonator.
	The filter's operation relies on equal emission of the dressed-qubit mode into the feedline at two spatially separate points such that they destructively interfere.
	This effect protects the qubit from decoherence, similar to the creation of subradiant states in giant atoms \cite{kockum_decoherence-free_2018,kannan_waveguide_2020}.
	
	The destructive interference effect can be qualitatively understood by decomposing the circuit in Fig.~\ref{fig:1}a into sub-circuits that include either only mutual capacitance (Fig.~\ref{fig:1}b) or mutual inductance (Fig.~\ref{fig:1}c).
	See Appendix~\ref{sec:app-ckt} for a detailed analysis of the circuit.
	We assume the feedline and transmission line resonator have a characteristic impedance of $Z_0$ and that the input and output ports are terminated with matched loads.
	The temporal phase of $V_1$ represents the phase of the dressed-qubit mode in the resonator (purple).
	The label $\angle\theta$ associated with $V_2$ and $V_3$ represents the phase $\theta$ relative to $V_1$.
	We assume a steady-state analysis of the form $e^{j\omega_q t+j\theta}$.
	
	First, we consider the sub-circuit with only mutual capacitance (Fig.~\ref{fig:1}b).
	From the capacitance current-voltage relation, the outgoing current through $C_m$ is shifted by $+90^\circ$ relative to $V_1$. 
	Due to the resistive termination, this current is in phase with $V_2$; $V_2$ thus is also shifted by $+90^\circ$ relative to $V_1$.
	Due to the $\lambda_q/4$ separation, $V_3$ is shifted by $-90^\circ$ relative to $V_2$; in total, $V_3$ is in phase with $V_1$.
	Second, we consider the sub-circuit with only mutual inductance (Fig.~\ref{fig:1}c).
	From the in-phase dot convention, $V_1$ is in-phase with $V_A$ and out-of-phase with $V_3$.
	Due to the $\lambda_q/4$ separation, $V_2$ is shifted by $-90^\circ$ relative to both $V_A$ and $V_1$.
	Finally, superimposing these phase relations in the full circuit in Fig.~\ref{fig:1}a, we see that the voltage contributions from the capacitive (blue) and inductive (red) couplings are out-of-phase and destructively interfere both to the left (solid lines) and the right (dashed lines).
	
	Maximizing the destructive interference in the feedline (i.e., matching the amplitudes of the out-of-phase voltages at the target qubit frequency $\wqt$) requires careful design of the ratio of mutual inductance to capacitance $r=L_m/C_m$. 
	We can derive an analytic expression for the optimal ratio $\ropt$ as (see Appendix~\ref{sec:app-ckt})
	\begin{equation}
		\begin{aligned}
			\ropt\approx Z_0^2\sin\left(\frac{2\pi}{\omega_r}\frac{\wqt}{4}\right).\\
		\end{aligned}
		\label{eq:ropt}
	\end{equation}
	For a target qubit frequency $\wqt/2\pi=\SI{5}{GHz}$, resonator frequency $\omega_r/2\pi=\SI{7}{GHz}$, and characteristic impedance $Z_0=\SI{50}{\ohm}$, we expect an optimal ratio of $L_m/C_m\approx\SI{2.25}{pH/fF}$.
	We note that the resonator linewidth is independent of $\ropt$; to increase the linewidth, both $C_m$ and $L_m$ can be increased by the same proportion.
	
	\begin{figure}[t]
		\centering
		\includegraphics{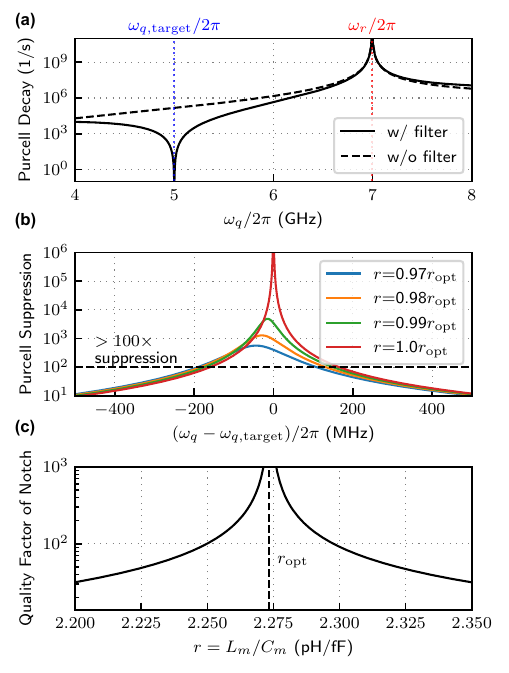}
		\caption{
			Circuit simulation. (a) Purcell decay with interferometric Purcell filter for optimal ratio $r=\ropt$. Unfiltered Purcell decay is shown for comparison.
			(b) Purcell suppression as a function of frequency for different ratios $r=L_m/C_m$. The horizontal dashed line provides a guide for the bandwidth of two orders of magnitude suppression.
			(c) The quality factor of bandstop response as a function of ratio $r$.  The vertical dashed line indicates the optimal ratio $\ropt$.
		} 
		\label{fig:2}
	\end{figure}
	
	We now perform a numerical analysis.
	The relaxation rate of a qubit to the environment is related to the input admittance from the perspective of the Josephson junction (JJ). 
	For a weakly anharmonic qubit such as the transmon, the Purcell decay closely matches the classical decay rate given by \cite{esteve_effect_1986,neeley_transformed_2008}
	\begin{equation}
		\Gp=\frac{\mathrm{Re}\left[Y_\text{in}\left(\omega_q\right)\right]}{C_\Sigma}\,,
		\label{eq:Gp_admittance}
	\end{equation}
	where $C_\Sigma$ is the total capacitance of the transmon island and $Y_\text{in}$ is the input admittance from the perspective of the JJ. 
	
	We compute $\Gp$ with our interferometric Purcell filter using circuit simulation of the admittance $Y_\text{in}(\omega_q)$ as shown in Fig.~\ref{fig:1}d, where we use a lumped model for the resonator for generality.
	To determine the optimal ratio $\ropt$, we keep $C_m$ fixed and sweep the inductive coupling $L_m$ (and thus the ratio $r$).
	The lumped inductor of the resonator is set accordingly to keep $\omega_r$ constant. 
	
	For circuit simulation, we assume a target qubit frequency of $\wqt/2\pi=\SI{5}{GHz}$, resonator frequency of $\omega_r/2\pi=\SI{7}{GHz}$, resonator linewidth of $\kappa/2\pi=\SI{12.3}{MHz}$, on-resonance resonator-qubit coupling strength of $g_0/2\pi=\SI{200}{MHz}$, and transmon charging energy of $E_C/h=\SI{200}{MHz}$.
	We note that the quoted $\kappa$ represents the effective resonator linewidth, i.e. the resonator linewidth in the presence of the Purcell filter.
	We assume a large  resonator linewidth as this has been shown to be beneficial for achieving fast qubit readout, but which comes at the cost of lower qubit lifetime in the absence of Purcell filtering \cite{sunada_fast_2022,walter_rapid_2017}.
	For the optimal ratio $\ropt$, we observe a sharp bandstop response in the Purcell decay rate, centered at $\wqt$ (see Fig.~\ref{fig:2}a).
	Using the same values for $\omega_r$, $g_0$, $\alpha$, and $\kappa$, the unfiltered Purcell decay rate is calculated as a function of $\omega_q$ using Eq.~\eqref{eq:Gp_unfiltered} and shown for comparison.
	
	The magnitude of Purcell suppression can be calculated as the ratio of the unfiltered to filtered Purcell decay rates, where $\omega_q$, $\omega_r$, $g_0$, $\alpha$, and $\kappa$ are held constant between the unfiltered and filtered cases; note, this implies that $\chi$ is also held constant, as $\chi= g^2\alpha/[(\omega_q-\omega_r)(\omega_q-\omega_r+\alpha)]$ \cite{koch_charge-insensitive_2007}.
	Since $\chi$ and $\kappa$ are held constant, the readout speed of the unfiltered and filtered cases can be expected to remain equivalent, assuming other parameters, such as photon number and measurement efficiency, are also fixed. 
	By this definition, the filter's achieved Purcell suppression can be fairly evaluated while maintaining the same readout speed as in the unfiltered case.
	
	We plot the Purcell suppression as a function of qubit frequency for different ratios $r$ in Fig.~\ref{fig:2}b.
	We define the bandstop quality factor $\Qf=\wqt/\Delta\omega$, where the notch is positioned at the target qubit frequency and $\Delta\omega$ is the full-width at half-minimum.
	Like a conventional bandstop filter component \cite{reed_fast_2010}, a larger quality factor $\Qf$ corresponds to a larger Purcell suppression at the notch frequency.
	We plot $\Qf$ as a function of the ratio $r=L_m/C_m$ in Fig.~\ref{fig:2}c.
	We see that the quality factor $\Qf$ is maximized near $\ropt\approx\SI{2.27}{pH/fF}$, in good agreement with our analytic estimate of $\SI{2.25}{pH/fF}$. 
	From Fig.~\ref{fig:2}b, we see that even though the quality factor $\Qf$ decreases for sub-optimal $r$, the bandwidth for which the Purcell suppression is more than two orders of magnitude remains roughly constant. 
	Finally, we note that the proposed design is readily extended to multi-qubit devices; see Appendix~\ref{sec:app-scaling} for further details.

	\section{Experimental Results}
	\begin{figure}[t]
		\centering
		\includegraphics{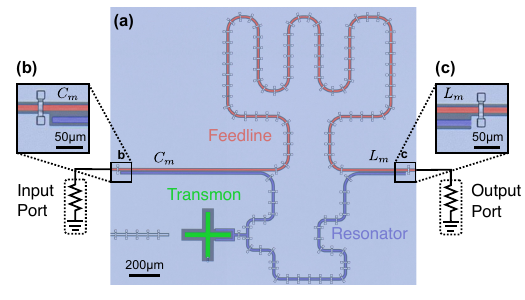}
		\caption{Experimental device. (a) False-colored micrograph of interferometric Purcell filter with \SI{50}{\ohm} feedline (red), transmon qubit (green), and quarter-wavelength resonator (blue) coupled to feedline (b) capacitively on the open end and (c) inductively on the shorted end.}
		\label{fig:micrograph}
	\end{figure}

	We experimentally demonstrate interferometric Purcell suppression with a device featuring a flux-tunable transmon qubit coupled to a $\lambda_r/4$ coplanar waveguide resonator. 
	An optical micrograph of the chip is shown in Fig.~\ref{fig:micrograph}a. All coplanar waveguides have a width of \SI{10}{\micro m} and a gap of \SI{6}{\micro m}. 
	The open end of the $\lambda_r/4$ resonator capacitively couples to the feedline (Fig.~\ref{fig:micrograph}b) with a center-to-center pitch of \SI{22}{\micro m} and \SI{1000}{\micro m} length. 
	The shorted end inductively couples to the feedline (Fig.~\ref{fig:micrograph}c) with a center-to-center pitch of \SI{24}{\micro m} and \SI{300}{\micro m} length. 
	A strong coupling is achieved with a claw-like structure coupled to a transmon qubit. 
	See Appendix~\ref{sec:app-setup} for details on the full experimental setup.
	When the qubit is biased at $\omega_q/2\pi=\SI{5330}{MHz}$, the fabricated readout resonator is measured to have a frequency of $\omega_r/2\pi=\SI{7578.5}{MHz}$ with resonator linewidth of $\kappa/2\pi=\SI{13.8}{MHz}$, as extracted from the resonator's transmission spectrum (see Appendix~\ref{sec:app-setup}).
	The transmon qubit is measured to have an anharmonicity of $\alpha/2\pi=\SI{-202}{MHz}$ and a dispersive shift of $\chi/2\pi=\SI{-1.6}{MHz}$.
	Using the standard dispersive shift approximation for a transmon qubit \cite{koch_charge-insensitive_2007}, we calculate a coupling strength of $g/2\pi=\SI{206}{MHz}$ at this qubit frequency.
	This corresponds to an on-resonance coupling strength of $g_0/2\pi=\SI{246}{MHz}$.
	
	We use the following procedure to experimentally extract the Purcell decay rate $\Gp$ of the qubit into the waveguide, following \cite{sunada_fast_2022}. By driving the qubit on resonance through the feedline with drive power $P$ and observing oscillations with Rabi rate $\Omega$, we can infer the Purcell decay rate into the feedline as given by
	\begin{equation}
		\Gp = \frac{\Omega^2}{2}\frac{\hbar\omega_q}{P}\,.
		\label{eq:Gp_Rabi}
	\end{equation}
	Using the flux-tunable transmon qubit, we can repeat this experiment and measure $\Omega$ across a wide range of qubit frequencies $\omega_q$.
	We note that Eq.~\eqref{eq:Gp_Rabi} differs from that in \cite{sunada_fast_2022} by a factor of 2 due to differences in the input-output theory of a resonator coupled to a two-sided versus a one-sided waveguide.
	See Appendix~\ref{sec:app-twosided} for further details.

\begin{figure}[t]
		\centering
		\includegraphics{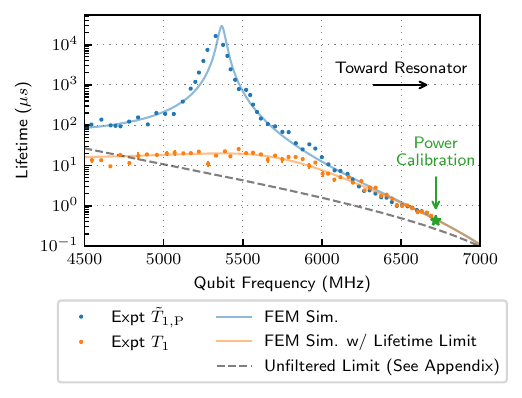}
		\caption{Experimental performance. The measured qubit lifetime $T_1$ and lower bound of the Purcell-limited lifetime $\Tpt$ are plotted as markers. Predicted $\Tp$ from FEM simulation with and without intrinsic qubit decay of \SI{20}{\micro s} are plotted as solid lines. A hypothetical lifetime limit without a Purcell filter is included for comparison. Power calibration is performed at \SI{6719}{MHz}. Error bars are shown for $T_1$ measurements if larger than the marker.}
		\label{fig:expt-data}
	\end{figure}

	To determine the applied drive power $P$, we use the following procedure to measure the attenuation between the room temperature equipment and the device.
	We first tune the qubit to a point near the resonator where we expect the measured qubit lifetime $T_1$ to be Purcell-limited.
	For instance, at $\omega_q/2\pi=\SI{6719}{MHz}$, we measure $T_1=\SI{0.4}{\micro s}$. 
	Because this lifetime is two orders of magnitude lower than the expected intrinsic qubit lifetime on the order of tens of microseconds, we make the assumption that Purcell decay is the dominant decay channel at this qubit frequency.
	Under this assumption, the Purcell-limited lifetime is equivalent to the measured relaxation time, i.e., $\Tp=1/\Gp\approx T_1$.
	For a Purcell-limited frequency point such as $\omega_q/2\pi=\SI{6719}{MHz}$, we can rewrite Eq.~\eqref{eq:Gp_Rabi} and calculate the applied drive power $P\approx \Omega^2\hbar\omega_qT_1/2$.
	We denote the power calibration point in green in Fig.~\ref{fig:expt-data}.
	
	While applicable at higher frequencies, this power calibration procedure cannot be used at lower frequencies near the notch frequency, since the qubit lifetime is not Purcell-limited.
	Instead, we choose to keep the extracted attenuation from the high-frequency point fixed for subsequent measurements.
	Since lower frequencies have less cable loss, the actual applied drive power will be higher.
	From Eq.~\eqref{eq:Gp_Rabi}, this results in an upper bound estimate of $\Gp$, or equivalently, a lower bound estimate of $\Tp$.
	We henceforth refer to this lower bound as $\Tpt$, representing a conservative estimate of the Purcell-limited lifetime of our device.
	
	We measure the lower bound of the Purcell-limited lifetime $\Tpt$ for qubit frequencies from 4500 to \SI{6700}{MHz}. 
	The measured $\Tpt$ and $T_1$ are plotted in Fig.~\ref{fig:expt-data}. 
	We note that the power calibration procedure is validated by the strong agreement between experimental $\Tpt$ and $T_1$ at higher qubit frequencies, where the qubit lifetime is Purcell-limited.
	The notch of the bandstop response is located at \SI{5330}{MHz}. 
	Relative to the unfiltered case given by Eq.~\eqref{eq:Gp_unfiltered} (see Appendix~\ref{sec:app-unfiltered}), our filter achieves Purcell suppression of more than 2 orders of magnitude over a bandwidth of $\SI{400}{MHz}$. 
	The experimental data indicates a quality factor of about 50 for the notch filter; using Fig.~\ref{fig:2}c, we estimate that the inductive to capacitive coupling ratio is within 2\% of $\ropt$.
	When the qubit is positioned at the notch frequency, the measured $\Tpt$ is more than \SI{16}{ms}, corresponding to a Purcell suppression factor of more than two thousand. 
	
	We predict the performance of our device using finite-element method (FEM) simulation in Ansys High Frequency Simulation Software (HFSS).  
	We model our device as a 3-port device with input and output ports and replace the Josephson junction with the third port. We then use Eq.~\eqref{eq:Gp_admittance} to find the expected Purcell-limited lifetime. 
	
	The total capacitance is simulated in Ansys Q3D.
	We extract the input admittance $Y_\text{in}$ from the scattering parameters of a driven modal simulation in HFSS and fine tune the permittivity to match the experimental notch frequency.
	This fine tuning of about a percent corrects for slight inaccuracies in assumptions of our model such as uniform airbridge heights and negligible oxidation and surface roughness.
	The Purcell-limited lifetime using our FEM model is plotted in Fig.~\ref{fig:expt-data} (blue) and demonstrates good agreement with measured $\Tpt$.
	Including an estimated lifetime limit of \SI{20}{\micro s} in our model (orange), we observe good agreement with measured $T_1$.
	This limit is within our expectation given the intrinsic decay of the qubit and decay through the dedicated drive line.

	\section{Conclusions and Outlook}
	In conclusion, we have demonstrated a technique to suppress the Purcell decay of a qubit using destructive interference.
	The interferometric Purcell filter  does not require dedicated filter components or impedance mismatch in the feedline, making it an attractive option for applications such as all-pass readout.
	We have presented an intuitive picture of the operation and design of this filter with circuit simulation.
	We have designed and fabricated a device with this filter and observed strong agreement between simulation and experiment.
	
	The use of interference to suppress Purcell decay is a versatile principle that can be extended to other lengths of transmission line resonators and for other combinations of couplings. 
	For example, the same effect can be achieved by coupling a $\lambda_r/2$ resonator capacitively at both ends to the waveguide at a separation of $\lambda_q/2$. 
	
	To minimize the footprint, we have elected to use a $\lambda_r/4$ resonator with $\lambda_q/4$ separation in the feedline.
	This makes the overall footprint of our device comparable to that of a conventional quarter-wavelength low-Q bandpass filter.
	In our presented layout, the capacitive coupling length takes up significant space; this could be compacted straightforwardly using meandering or interdigitated capacitors \cite{steffen_high-coherence_2010}.
	
	\section*{Acknowledgements}
	The authors thank Neereja M. Sundaresan, Oliver E. Dial, and David W. Abraham for insightful discussions.
	
	This work was supported in part by the MIT-IBM Watson AI Lab. This material is based upon work supported by the Under Secretary of Defense for Research and Engineering under Air Force Contract No. FA8702-15-D-0001. Any opinions, findings, conclusions or recommendations expressed in this material are those of the author(s) and do not necessarily reflect the views of the Under Secretary of Defense for Research and Engineering. A.Y. and J.W. acknowledge support from the NSF Graduate Research Fellowship. Y.Y. acknowledges support from the IBM PhD Fellowship and the NSERC Postgraduate Scholarship. G.C. acknowledges support from the Harvard Graduate School of Arts and Sciences Prize Fellowship.
	
	\appendix
	
	\section{Unfiltered Purcell Limit}
	\label{sec:app-unfiltered}
	\begin{figure}
		\includegraphics{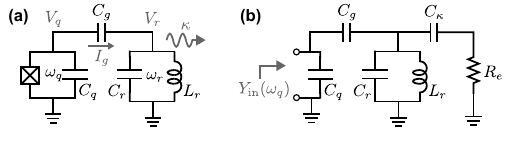}
		\caption{(a) Circuit for the analytic derivation of unfiltered Purcell limit. (b) Circuit for numerical (SPICE) simulation.}
		\label{fig:app-unfiltered}
	\end{figure}
	
	\begin{figure}
		\includegraphics{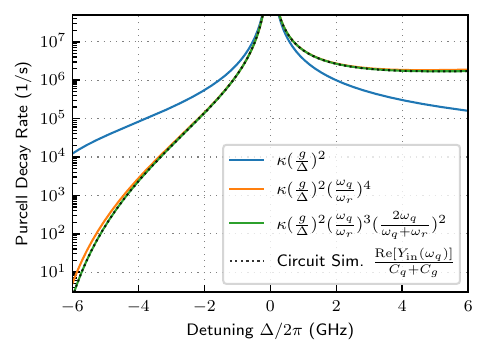}
		\caption{Comparison of analytic and numerical (SPICE) Purcell decay rates for a qubit coupled to a lossy single-mode resonator, assuming $\omega_r=\SI{7}{GHz}$. The exact (green) and approximate (orange) expressions demonstrate excellent agreement with numerical results. The standard expression (blue) overestimates Purcell decay for negative detunings. Note that $g$ has frequency dependence, given by $g=g_0\sqrt{\omega_q/\omega_r}$.}
		\label{fig:app-spice}
	\end{figure}
	
	We derive an analytic expression for the Purcell limit of a qubit capacitively coupled to a lossy single-mode resonator, as in Fig.~\ref{fig:app-unfiltered}a (see Appendix~\ref{sec:app-analytic}).
	We compare with numerical  simulations of the circuit in Fig.~\ref{fig:app-unfiltered}b (see Appendix~\ref{sec:app-spice}) using the classical decay time expression from Eq.~\eqref{eq:Gp_admittance}.
	This classical treatment has been found to be a valid approximation for weakly anharmonic qubits such as the transmon qubit \cite{houck_controlling_2008,sunada_fast_2022,reed_fast_2010,bronn_broadband_2015}.
	As such, our analytic derivation treats the qubit as a linear resonator.
	
	First, we summarize the key results.
	We find the unfiltered Purcell decay is given by
	\begin{equation}
		\begin{aligned}
			\Gp&=\kappa\left(\frac{g}{\Delta}\right)^2\left(\frac{\omega_q}{\omega_r}\right)^3\left(\frac{2\omega_q}{\omega_q+\omega_r}\right)^2\\
			&\approx\kappa\left(\frac{g}{\Delta}\right)^2\left(\frac{\omega_q}{\omega_r}\right)^4,
		\end{aligned}
		\label{eq:Gp_exact}
	\end{equation}
	where $\Delta=\omega_q-\omega_r$.
	The addition of the $\sim\!\!(\omega_q/\omega_r)^4$ factor is in significant contrast to the standard expression of
	\begin{equation}
		\Gp=\kappa\left(\frac{g}{\Delta}\right)^2.
		\label{eq:Gp_std}
	\end{equation}
	We note that both \eqref{eq:Gp_exact} and \eqref{eq:Gp_std}  assume frequency dependence of $g=g_0\sqrt{\omega_q/\omega_r}$, as expected from capacitive coupling.
	To compare these expressions with numerics, we perform a SPICE simulation of the circuit shown in Fig.~\ref{fig:app-unfiltered}b.
	We assume parameters of $\kappa/2\pi=\SI{12.3}{MHz}$, $\omega_r/2\pi=\SI{7}{GHz}$, and $g_0/2\pi=\SI{200}{MHz}$.
	The Purcell decay is calculated using the admittance calculation in \eqref{eq:Gp_admittance}, demonstrating excellent agreement with the analytic expression of \eqref{eq:Gp_exact}, as shown in Fig.~\ref{fig:app-spice}.
	We see that for $\omega_q<\omega_r$, the standard expression overestimates the unfiltered Purcell decay.
	We consider the implications of this on estimates of Purcell suppression by filters reported in the literature, which use the standard Purcell decay approximation in \eqref{eq:Gp_std}.
	For typical negative detunings of \SI{-1.5}{GHz} and \SI{-2.5}{GHz}, usage of the standard expression in \eqref{eq:Gp_std} would overestimate Purcell suppression by a factor of $2.7\times$ and $6.1\times$, respectively.
	Although our analysis does not include the multi-mode effects of the resonator, our presented results still  hold for negative detunings where $\omega_q<\omega_r$ \cite{houck_controlling_2008}.
	
	\subsection{Analytic Derivation}
	\label{sec:app-analytic}
	We derive \eqref{eq:Gp_exact}, the unfiltered Purcell decay rate, in a similar manner to that which has been done for a qubit dispersively coupled to a readout resonator with a bandpass Purcell filter \cite{jeffrey_fast_2014}.
	The quality factor $Q_q$ of the qubit is given by
	\begin{equation}
		Q_q\equiv \frac{\text{energy stored in qubit}}{\text{energy lost / qubit cycle}}\,.
		\label{eq:app-Qq}
	\end{equation}
	The qubit's stored energy is
	\begin{equation}
		E_q = \frac{1}{2}C_qV_q^2\,.
	\end{equation}
	Assuming the resonator is the only lossy element, the energy lost per resonator cycle is given by
	\begin{equation}
		\text{energy lost / resonator cycle} = \frac{E_r}{Q_r} = \frac{\frac{1}{2}C_rV_r^2}{Q_r}\,.
	\end{equation}
	Since the qubit and resonator are detuned, there is a correction factor of
	\begin{equation}
		\frac{\text{energy lost / qubit cycle}}{\text{energy lost / resonator cycle}}=\frac{\text{resonator cycle}}{\text{qubit cycle}} = \frac{\omega_q}{\omega_r}.
	\end{equation}
	Then \eqref{eq:app-Qq} becomes
	\begin{equation}
		Q_q=Q_r\frac{C_q}{C_r}\frac{\abs{V_q}^2}{\abs{V_r}^2}\frac{\omega_r}{\omega_q}\,.
		\label{eq:app-Qq2}
	\end{equation}
	We analyze the circuit at the qubit frequency (i.e., if frequency-dependent, circuit parameters such as impedance are assumed to be at $\omega_q$).
	Because the qubit and resonator modes are detuned, we can assume that the impedance of the resonator is smaller than that of the coupling capacitor $C_g$, i.e., $Z_r\ll Z_g$.
	In this case, then the current through the capacitance is given by $I_g= V_q/Z_g$.
	Then we can obtain $V_r=I_gZ_r= V_qZ_r/Z_g$ and thus
	\begin{equation}
		\frac{V_q}{V_r}=\frac{Z_g}{Z_r}\,\,.
	\end{equation}
	Substituting into \eqref{eq:app-Qq2}, we obtain
	\begin{equation}
		Q_q= Q_r\frac{C_q}{C_r}\frac{\abs{Z_g}^2}{\abs{Z_r}^2}\frac{\omega_r}{\omega_q}\,.
		\label{eq:app-Qq3}
	\end{equation}
	Substituting $|Z_g|^2=1/\omega_q^2C_g^2$, $\kappa=\omega_r/Q_r$, and $\Gp=\omega_q/Q_q$, this simplifies to
	\begin{equation}
		\Gp=\kappa\frac{C_r}{C_q}{|Z_r|}^2\omega_q^2C_g^2\left(\frac{\omega_q}{\omega_r}\right)^2.
	\end{equation}
	The impedance of the resonator at $\omega_q$ is given by
	\begin{equation}
		\begin{aligned}
			Z_r&=\frac{1}{\frac{1}{j\omega_qL_r}+j\omega_qC_r}
			=\frac{Z^0_r}{j}\frac{\omega_q\omega_r}{\Delta(\omega_q+\omega_r)}\,.
		\end{aligned}
	\end{equation}
	Here, $Z^0_r=\sqrt{L_r/C_r}$ is the impedance of the resonator, and $\omega_r=1/\sqrt{L_rC_r}$ is the resonator frequency, where we have made the assumption that $C_r\gg C_g$.
	Our expression then becomes
	\begin{equation}
		\Gp=\kappa\left(\frac{1}{\Delta}\right)^2\frac{C_r}{C_q}{Z^0_r}^2\omega_q^2C_g^2\left(\frac{\omega_q}{\omega_r}\right)^2	\left(\frac{\omega_q\omega_r}{\omega_q+\omega_r}\right)^2.
	\end{equation}
	Substituting ${Z_r^0}^2=1/\omega_r^2C_r^2$, this simplifies to
	\begin{equation}
		\Gp=\kappa\left(\frac{1}{\Delta}\right)^2\frac{C_g^2\omega_r^2}{C_qC_r}\left(\frac{\omega_q}{\omega_r}\right)^4	\left(\frac{\omega_q}{\omega_q+\omega_r}\right)^2.
	\end{equation}
	Assuming that $C_q,C_r\gg C_g$, the standard qubit-resonator capacitive coupling is then given by
	\begin{equation}
		\begin{aligned}
			g&=\frac{1}{2}\frac{C_g}{\sqrt{C_qC_r}}\sqrt{\omega_q\omega_r}=\frac{1}{2}\frac{C_g\omega_r}{\sqrt{C_qC_r}}\sqrt{\frac{\omega_q}{\omega_r}}=g_0\sqrt{\frac{\omega_q}{\omega_r}}.
		\end{aligned}
	\end{equation}
	We thus arrive at our final expression of
	\begin{equation}
		\Gp=\kappa\left(\frac{g}{\Delta}\right)^2\left(\frac{\omega_q}{\omega_r}\right)^3	\left(\frac{2\omega_q}{\omega_q+\omega_r}\right)^2.
		\label{eq:exact-nofilter}
	\end{equation}
	The last factor can be approximated with a simpler form. Defining $x=\Delta/\omega_r=(\omega_q-\omega_r)/\omega_r$, the last factor can be rewritten as 
	\begin{equation}
		f(x)=\left(\frac{2\omega_q}{\omega_q+\omega_r}\right)^2=4\left(\frac{x+1}{x+2}\right)^2.
	\end{equation}
	Performing a Taylor expansion to first order about $x=0$ (i.e., the zero detuning point), we find that $f(x)\approx f(0)+f'(0)x=1+x=\omega_q/\omega_r$.
	Thus, \eqref{eq:exact-nofilter} is well approximated by
	\begin{equation}
		\Gp\approx\kappa\left(\frac{g}{\Delta}\right)^2\left(\frac{\omega_q}{\omega_r}\right)^4.
		\label{eq:approx-nofilter}
	\end{equation}
	
	\subsection{Numerical Simulation}
	\label{sec:app-spice}
	We obtain the Purcell decay using numerical simulation of the circuit in SPICE.
	The circuit is shown in Fig.~\ref{fig:app-unfiltered}b.
	The resonator frequency $\omega_r$ is given by
	\begin{equation}
		\omega_r=\frac{1}{\sqrt{L_r(C_r+C_\kappa+C_g)}}\,.
	\end{equation}
	The coupling rate $g$ is given by
	\begin{equation}
		g=\frac{1}{2}\frac{C_g}{\sqrt{(C_q+C_g)(C_r+C_\kappa+C_g)}}\sqrt{\omega_q\omega_r}\,.
	\end{equation}
	The decay rate $\kappa$ is given by
	\begin{equation}
		\kappa=\frac{R_e}{L_r}\left(\frac{C_\kappa}{C_r+C_\kappa+C_g}\right)^2.
	\end{equation}
	We choose circuit parameters that satisfy the chosen parameters $\omega_r$, $g$, and $\kappa$. 
	Using \eqref{eq:Gp_admittance}, the Purcell decay rate is given by
	\begin{equation}
		\Gp=\frac{\mathrm{Re}\left[Y_\text{in}\left(\omega_q\right)\right]}{C_q+C_g},
	\end{equation}
	where $Y_\text{in}$ is the admittance from the perspective of the Josephson junction.
	We confirm good agreement between these analytic expressions and numerical simulation.
	
	\section{\texorpdfstring{Analytic Derivation of Optimal $\mathbf{L_m/C_m}$}{Analytic Derivation of Optimal Mutual Inductance to Capacitance Ratio}}
	\label{sec:app-ckt}
	
	\begin{figure}[t]
		\centering
		\includegraphics{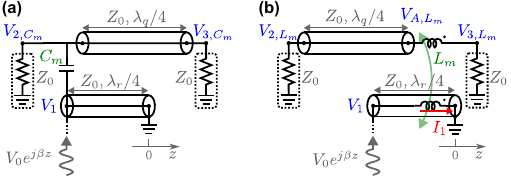}
		\caption{Sub-circuits of either only (a) mutual capacitance or (b) mutual inductance.}
		\label{fig:app-ckt}
	\end{figure}
	Here, we derive the ratio $r=L_m/C_m$ that maximizes the destructive interference of the qubit mode in the feedline, as described in the main text.
	We perform frequency domain analysis at the target qubit frequency $\wqt$ and assume the sinusoidal steady state condition.
	In other words, presented values of voltage (e.g., $V_i$) and current (e.g., $I_i$) are assumed to have time dependence of the form $e^{j\omega_q t}$.
	Complex values (e.g., $e^{j\theta}$) indicate the temporal phase $\theta$.
	
	First, we find the voltage and current in the quarter-wavelength resonator (see Fig.~\ref{fig:app-ckt}a-b).
	We assume some incident wave at $V_1$ at the target qubit frequency $\wqt$, described by $V_0e^{j\beta z}$ where $\beta=2\pi/\lambda_q$. 
	A transmission line with characteristic impedance $Z_0$ terminated in a short has voltage and current spatial dependence given by \cite{pozar_microwave_2012}
	\begin{equation}
		V_r(z)=-2jV_0\sin\left(\beta z\right)
	\end{equation}
	and
	\begin{equation}
		I_r(z)=\frac{2V_0}{Z_0}\cos\left(\beta z\right),
	\end{equation}
	where $z$ is the position with respect to the short termination.
	The voltage at the open end is given by
	\begin{equation}
		\begin{aligned}
			V_1=V_r(z=-\frac{\lambda_r}{4})&=2jV_0\sin\left(\frac{2\pi}{\lambda_q}\frac{\lambda_r}{4}\right)\\
			&=2jV_0\sin\left(\frac{2\pi}{\omega_r}\frac{\wqt}{4}\right).
		\end{aligned}
		\label{eq:app-V1}
	\end{equation}
	Note that this reflects that the mode with a wavelength of $\lambda_q$ has an antinode that does not align with the open of the $\lambda_r/4$ resonator.
	The current at the shorted end is given by
	\begin{equation}
		I_1=I_r(z=0)=\frac{2V_0}{Z_0}\,.
		\label{eq:app-I1}
	\end{equation}
	
	Now, we consider the circuit with only mutual capacitance (Fig.~\ref{fig:app-ckt}a) and calculate the contributions to $V_2$ and $V_3$, which we denote as $V_{2,C_m}$ and $V_{3,C_m}$, respectively.
	We assume the mutual capacitance is small in magnitude such that $Z_0\omega C_m\ll 1$.
	The voltage $V_{2,C_m}$ can then be approximated by voltage division as
	\begin{equation}
		V_{2,C_m}=\frac{\frac{Z_0}{2}}{\frac{Z_0}{2}+\frac{1}{j\wqt C_m}}V_1
		\approx \frac{1}{2}j\wqt C_m Z_0 V_1.
	\end{equation}
	Substituting in \eqref{eq:app-V1}, this becomes
	\begin{equation}
		V_{2,C_m} \approx -\wqt C_m Z_0V_0\sin\left(\frac{2\pi}{\omega_r}\frac{\wqt}{4}\right).
		\label{eq:app-V2Cm}
	\end{equation}
	The $\lambda_q/4$ separation introduces a phase delay such that \cite{pozar_microwave_2012}
	\begin{equation}
		V_{3,C_m}=e^{-j\beta l}V_{2,C_m}=e^{-j\frac{2\pi}{\lambda_q}\frac{\lambda_q}{4}}V_{2,C_m}=-jV_{2,C_m}.
		\label{eq:app-delay}
	\end{equation}
	The voltage $V_{3,C_m}$ is then given by 
	\begin{equation}
		V_{3,C_m}\approx j\wqt C_m Z_0V_0\sin\left(\frac{2\pi}{\omega_r}\frac{\wqt}{4}\right).
		\label{eq:app-V3Cm}
	\end{equation}
	
	Next, we derive the contributions of $V_{2,L_m}$ and $V_{3,L_m}$ by the mutual inductance $L_m$, considering the circuit in Fig.~\ref{fig:app-ckt}b. 
	In this simple model, we assume that the coupling coefficient is $k=1$.
	Because $I_1$ is at the shorted end of the resonator, we can infer that $I_1$ is much larger than the current in the feedline, which has $Z_0$ terminations on either side.
	The voltage across the lumped inductance in the feedline is then approximated by
	\begin{equation}
		V_{A,L_m}-V_{3,L_m}\approx j\wqt L_mI_1.
	\end{equation}
	By Kirchoff's current law, $V_{3,L_m}/Z_0=-V_{A,L_m}/Z_0$.
	Then, we find
	\begin{equation}
		V_{3,L_m}\approx -\frac{1}{2}j\wqt L_mI_1=-j\wqt L_m\frac{V_0}{Z_0},
		\label{eq:app-V3Lm}
	\end{equation}
	where we have substituted in \eqref{eq:app-I1}.
	Similar to \eqref{eq:app-delay}, the voltage $V_{2,L_m}$ is given by
	\begin{equation}
		V_{2,L_m}=-jV_{A,L_m}=jV_{3,L_m}\approx\wqt L_m\frac{V_0}{Z_0}.
		\label{eq:app-V2Lm}
	\end{equation}
	
	We can now consider the superposition of these two sub-circuits. 
	We can derive the ratio of mutual inductance to capacitance that will maximize the destructive interference at $V_2$ and $V_3$.
	From \eqref{eq:app-V2Cm} and \eqref{eq:app-V3Cm}, we see that $V_2=V_{2,C_m}+V_{2,L_m}=0$ if
	\begin{equation}
		\begin{aligned}
			\wqt L_m\frac{V_0}{Z_0}&=\wqt C_m Z_0V_0\sin\left(\frac{2\pi}{\omega_r}\frac{\wqt}{4}\right)\\
			\frac{L_m}{C_m}&=Z_0^2\sin\left(\frac{2\pi}{\omega_r}\frac{\wqt}{4}\right).\\
		\end{aligned}
	\end{equation}
	From \eqref{eq:app-V3Lm} and \eqref{eq:app-V2Lm}, we see that this is also the condition such that $V_3=V_{3,C_m}+V_{3,L_m}=0$.
	We refer to this as the optimal ratio $\ropt$ in \eqref{eq:ropt} in the main text.
	
	\section{Scaling to Multiple Qubits}
	\label{sec:app-scaling}
	\begin{figure}[t]
		\centering
		\includegraphics{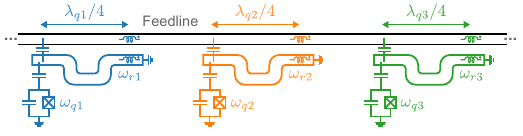}
		\caption{
			Illustration of interferometric Purcell suppression applied to multiplexed qubit readout.
		}
		\label{fig:app-cartoon}
	\end{figure}
	The proposed scheme can readily be extended to multi-qubit devices, such as for multiplexed qubit readout \cite{heinsoo_rapid_2018}. 
	Because none of the readout resonators or qubits share the same frequency, each readout unit (consisting of a resonator and qubit) can be designed independently and coupled to the same feedline.
	An illustration of a multiplexed qubit readout circuit with interferometric Purcell suppression is shown in Fig.~\ref{fig:app-cartoon}.
	In this illustration, each qubit's frequency is assumed to be higher than that of its corresponding resonator, as each quarter-wavelength resonator is longer than its corresponding $\lambda_{qi}/4$ separation in the feedline.

	\section{Sample and Setup }
	\label{sec:app-setup}
	\begin{figure}[t]
		\centering
		\includegraphics{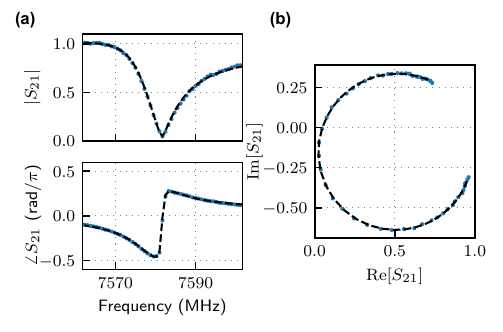}
		\caption{Transmission spectrum of resonator and fit using the method described in \cite{probst_efficient_2015}. (a)~Magnitude and phase versus frequency. (b)~Complex plane. }
		\label{fig:app-fit}
	\end{figure}
	\begin{figure}
		\centering
		\includegraphics{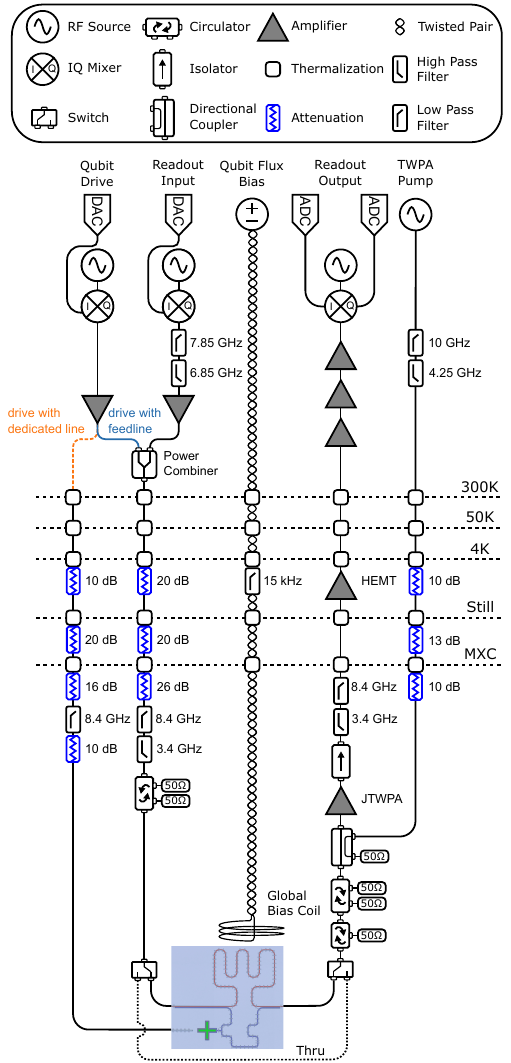} 
		\caption{Experimental diagram and wiring.} 
		\label{fig:fridge-diagram}
	\end{figure}
	The flux-tunable transmon qubit and transmission line resonator are comprised of layers of thin-film aluminum on a silicon substrate.
	The diagram of the experimental setup is shown in Fig.~\ref{fig:fridge-diagram}.
	We test the device in a Bluefors LD400 dilution refrigerator at a base temperature of \SI{20}{mK} at the mixing chamber (MXC).
	The device is housed inside a superconducting aluminum shield, nested within a Cryoperm shield mounted at the MXC.
	Microwave control is applied using the QICK ZCU111 RFSoC FPGA \cite{stefanazzi_qick_2022}. 
	
	As shown in Fig.~\ref{fig:fridge-diagram}, to drive the qubit through the feedline, the qubit and readout drives are combined with a power combiner (see solid blue line). For other measurements, the qubit drive is connected to a dedicated drive line (see dotted orange line).
	
	To determine the resonator linewidth, we measure the transmission of our resonator, as shown in Fig.~\ref{fig:app-fit}. 
	The measurement is calibrated to a through line using a switch bank (see Fig.~\ref{fig:fridge-diagram}).
	We extract the linewidth of our resonator using the techniques detailed in \cite{probst_efficient_2015} and extract fit parameters of $\omega_r/2\pi=\SI{7578.5}{MHz}$, $\kappa/2\pi=\SI{13.8}{MHz}$, and internal quality factor of $Q_i=\SI{9.6e3}{}$.

	\section{Decay of a Qubit Mode into a Waveguide}
	\label{sec:app-twosided}
	\begin{figure}[t!]
		\centering
		\includegraphics{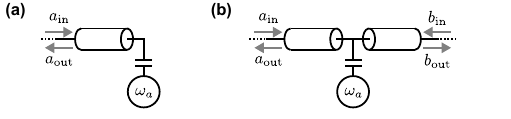}
		\caption{Input-output networks for a mode with coupling to (a)~one-sided and (b)~two-sided waveguides.}
		\label{fig:app-cmt}
	\end{figure}
	Here, we detail the origin of the factor of 2 difference of Eq.~\eqref{eq:Gp_Rabi} compared to that in Ref.~\cite{sunada_fast_2022}, where a one-sided waveguide is used.
	Input-output networks for a qubit mode coupled to a one-sided and two-sided waveguide are shown in Fig.~\ref{fig:app-cmt}a-b.
	
	\subsection{One-Sided Waveguide}
	Following the derivation of \cite{blais_circuit_2021}, the equation of motion for a qubit mode $\ah$ coupled to a one-sided waveguide (Fig.~\ref{fig:app-cmt}a) is given by 
	\begin{equation}
		\frac{d\ah}{dt}=\frac{i}{\hbar}[\hat{H}_q,\ah]-\frac{\gam}{2}\ah+\sqrt{\gam}\ain(t)\,,
	\end{equation}
	where $\hat{H}_q$ is the Hamiltonian of the qubit, $\ain(t)$ is a time-varying input field and $\gam$ is the decay rate of the mode into the environment.
	In the one-sided waveguide case, this is equal to the rate at which the input drive excites the mode.
	
	To include a coherent classical drive, the input field undergoes a displacement transformation of $\ain(t)\to \ain(t)+A\exp(-i(\omega_d t+\phi_d))$. Equivalently, this extra term can be absorbed into the Hamiltonian as $\hat{H}_q\to \hat{H}_q+\hat{H}_d$ where the driving Hamiltonian is given by
	\begin{equation}
		\hat{H}_d=i\hbar\epsilon\left(\ah^\dagger e^{-i(\omega_dt+\phi_d)}+\ah e^{i(\omega_dt+\phi_d)}\right),
		\label{eq:app-Hdrive}
	\end{equation}
	where $\epsilon=A\sqrt{\gam}$. Since the Rabi frequency is given by $\Omega=2\epsilon$, it follows that
	\begin{equation}
		\frac{\Omega^2}{4}=A^2\gam\,.
	\end{equation}
	Finally, we convert from drive amplitude $A$ to input power using the relations of photon flux $\ndot=A^2$ and input power $\Pin=\hbar\omega_d\ndot$.
	We then obtain
	\begin{equation}
		\gam=\frac{\Omega^2}{4} \frac{\hbar\omega_d}{\Pin}\,,
	\end{equation}
	which matches the expression of \cite{sunada_fast_2022}.
	
	\subsection{Two-Sided Waveguide}
	The key distinguishing feature of a two-sided waveguide (Fig.~\ref{fig:app-cmt}b) is found in the equation of motion given by 
	\begin{equation}
		\frac{d\ah}{dt}=\frac{i}{\hbar}[\hat{H}_q,\ah]-\frac{\gam}{2}\ah+\sqrt{\frac{\gam}{2}}\left(\ain(t)+\bin(t)\right).
		\label{eq:EOM2}
	\end{equation}
	By energy conservation, the rate at which the mode can be excited by either $\ain(t)$ or $\bin(t)$ is half of the total decay rate $\gamma$ into the environment \cite{manolatou_coupling_1999}.
	Following the same procedure as before, we apply a coherent classical drive from the left so that $\ain(t)\to \ain(t)+A\exp(-i(\omega_d t+\phi_d))$.
	Like before, we can make the equivalent replacement of the Hamiltonian $\hat{H}_q\to \hat{H}_q+\hat{H}_d$ that corresponds to the same drive Hamiltonian as \eqref{eq:app-Hdrive}, but where $\epsilon=A\sqrt{\gam/2}$. 
	Rewriting in terms of Rabi frequency and input power, we find
	\begin{equation}
		\gam=\frac{\Omega^2}{2} \frac{\hbar\omega_d}{\Pin}\,,
	\end{equation}
	which matches Eq.~\eqref{eq:Gp_Rabi} from the main text, where we have driven the qubit on resonance $\omega_q=\omega_d$ and the decay rate of the qubit mode into the environment is the Purcell decay rate $\Gp=\gamma$.

	\bibliography{main.bib}

\begin{thebibliography}{34}%
\makeatletter
\providecommand \@ifxundefined [1]{%
 \@ifx{#1\undefined}
}%
\providecommand \@ifnum [1]{%
 \ifnum #1\expandafter \@firstoftwo
 \else \expandafter \@secondoftwo
 \fi
}%
\providecommand \@ifx [1]{%
 \ifx #1\expandafter \@firstoftwo
 \else \expandafter \@secondoftwo
 \fi
}%
\providecommand \natexlab [1]{#1}%
\providecommand \enquote  [1]{``#1''}%
\providecommand \bibnamefont  [1]{#1}%
\providecommand \bibfnamefont [1]{#1}%
\providecommand \citenamefont [1]{#1}%
\providecommand \href@noop [0]{\@secondoftwo}%
\providecommand \href [0]{\begingroup \@sanitize@url \@href}%
\providecommand \@href[1]{\@@startlink{#1}\@@href}%
\providecommand \@@href[1]{\endgroup#1\@@endlink}%
\providecommand \@sanitize@url [0]{\catcode `\\12\catcode `\$12\catcode
  `\&12\catcode `\#12\catcode `\^12\catcode `\_12\catcode `\%12\relax}%
\providecommand \@@startlink[1]{}%
\providecommand \@@endlink[0]{}%
\providecommand \url  [0]{\begingroup\@sanitize@url \@url }%
\providecommand \@url [1]{\endgroup\@href {#1}{\urlprefix }}%
\providecommand \urlprefix  [0]{URL }%
\providecommand \Eprint [0]{\href }%
\providecommand \doibase [0]{https://doi.org/}%
\providecommand \selectlanguage [0]{\@gobble}%
\providecommand \bibinfo  [0]{\@secondoftwo}%
\providecommand \bibfield  [0]{\@secondoftwo}%
\providecommand \translation [1]{[#1]}%
\providecommand \BibitemOpen [0]{}%
\providecommand \bibitemStop [0]{}%
\providecommand \bibitemNoStop [0]{.\EOS\space}%
\providecommand \EOS [0]{\spacefactor3000\relax}%
\providecommand \BibitemShut  [1]{\csname bibitem#1\endcsname}%
\let\auto@bib@innerbib\@empty
\bibitem [{\citenamefont {Blais}\ \emph {et~al.}(2004)\citenamefont {Blais},
  \citenamefont {Huang}, \citenamefont {Wallraff}, \citenamefont {Girvin},\
  and\ \citenamefont {Schoelkopf}}]{blais_cavity_2004}%
  \BibitemOpen
  \bibfield  {author} {\bibinfo {author} {\bibfnamefont {A.}~\bibnamefont
  {Blais}}, \bibinfo {author} {\bibfnamefont {R.-S.}\ \bibnamefont {Huang}},
  \bibinfo {author} {\bibfnamefont {A.}~\bibnamefont {Wallraff}}, \bibinfo
  {author} {\bibfnamefont {S.~M.}\ \bibnamefont {Girvin}},\ and\ \bibinfo
  {author} {\bibfnamefont {R.~J.}\ \bibnamefont {Schoelkopf}},\ }\bibfield
  {title} {\bibinfo {title} {Cavity quantum electrodynamics for superconducting
  electrical circuits: {{An}} architecture for quantum computation},\ }\href
  {https://doi.org/10.1103/PhysRevA.69.062320} {\bibfield  {journal} {\bibinfo
  {journal} {Physical Review A}\ }\textbf {\bibinfo {volume} {69}},\ \bibinfo
  {pages} {062320} (\bibinfo {year} {2004})}\BibitemShut {NoStop}%
\bibitem [{\citenamefont {Wallraff}\ \emph {et~al.}(2004)\citenamefont
  {Wallraff}, \citenamefont {Schuster}, \citenamefont {Blais}, \citenamefont
  {Frunzio}, \citenamefont {Huang}, \citenamefont {Majer}, \citenamefont
  {Kumar}, \citenamefont {Girvin},\ and\ \citenamefont
  {Schoelkopf}}]{wallraff_strong_2004}%
  \BibitemOpen
  \bibfield  {author} {\bibinfo {author} {\bibfnamefont {A.}~\bibnamefont
  {Wallraff}}, \bibinfo {author} {\bibfnamefont {D.~I.}\ \bibnamefont
  {Schuster}}, \bibinfo {author} {\bibfnamefont {A.}~\bibnamefont {Blais}},
  \bibinfo {author} {\bibfnamefont {L.}~\bibnamefont {Frunzio}}, \bibinfo
  {author} {\bibfnamefont {R.-S.}\ \bibnamefont {Huang}}, \bibinfo {author}
  {\bibfnamefont {J.}~\bibnamefont {Majer}}, \bibinfo {author} {\bibfnamefont
  {S.}~\bibnamefont {Kumar}}, \bibinfo {author} {\bibfnamefont {S.~M.}\
  \bibnamefont {Girvin}},\ and\ \bibinfo {author} {\bibfnamefont {R.~J.}\
  \bibnamefont {Schoelkopf}},\ }\bibfield  {title} {\bibinfo {title} {Strong
  coupling of a single photon to a superconducting qubit using circuit quantum
  electrodynamics},\ }\href {https://doi.org/10.1038/nature02851} {\bibfield
  {journal} {\bibinfo  {journal} {Nature}\ }\textbf {\bibinfo {volume} {431}},\
  \bibinfo {pages} {162} (\bibinfo {year} {2004})}\BibitemShut {NoStop}%
\bibitem [{\citenamefont {Wallraff}\ \emph {et~al.}(2005)\citenamefont
  {Wallraff}, \citenamefont {Schuster}, \citenamefont {Blais}, \citenamefont
  {Frunzio}, \citenamefont {Majer}, \citenamefont {Devoret}, \citenamefont
  {Girvin},\ and\ \citenamefont {Schoelkopf}}]{wallraff_approaching_2005}%
  \BibitemOpen
  \bibfield  {author} {\bibinfo {author} {\bibfnamefont {A.}~\bibnamefont
  {Wallraff}}, \bibinfo {author} {\bibfnamefont {D.~I.}\ \bibnamefont
  {Schuster}}, \bibinfo {author} {\bibfnamefont {A.}~\bibnamefont {Blais}},
  \bibinfo {author} {\bibfnamefont {L.}~\bibnamefont {Frunzio}}, \bibinfo
  {author} {\bibfnamefont {J.}~\bibnamefont {Majer}}, \bibinfo {author}
  {\bibfnamefont {M.~H.}\ \bibnamefont {Devoret}}, \bibinfo {author}
  {\bibfnamefont {S.~M.}\ \bibnamefont {Girvin}},\ and\ \bibinfo {author}
  {\bibfnamefont {R.~J.}\ \bibnamefont {Schoelkopf}},\ }\bibfield  {title}
  {\bibinfo {title} {Approaching {{Unit Visibility}} for {{Control}} of a
  {{Superconducting Qubit}} with {{Dispersive Readout}}},\ }\href
  {https://doi.org/10.1103/PhysRevLett.95.060501} {\bibfield  {journal}
  {\bibinfo  {journal} {Physical Review Letters}\ }\textbf {\bibinfo {volume}
  {95}},\ \bibinfo {pages} {060501} (\bibinfo {year} {2005})}\BibitemShut
  {NoStop}%
\bibitem [{\citenamefont {Gambetta}\ \emph {et~al.}(2006)\citenamefont
  {Gambetta}, \citenamefont {Blais}, \citenamefont {Schuster}, \citenamefont
  {Wallraff}, \citenamefont {Frunzio}, \citenamefont {Majer}, \citenamefont
  {Devoret}, \citenamefont {Girvin},\ and\ \citenamefont
  {Schoelkopf}}]{gambetta_qubit-photon_2006}%
  \BibitemOpen
  \bibfield  {author} {\bibinfo {author} {\bibfnamefont {J.}~\bibnamefont
  {Gambetta}}, \bibinfo {author} {\bibfnamefont {A.}~\bibnamefont {Blais}},
  \bibinfo {author} {\bibfnamefont {D.~I.}\ \bibnamefont {Schuster}}, \bibinfo
  {author} {\bibfnamefont {A.}~\bibnamefont {Wallraff}}, \bibinfo {author}
  {\bibfnamefont {L.}~\bibnamefont {Frunzio}}, \bibinfo {author} {\bibfnamefont
  {J.}~\bibnamefont {Majer}}, \bibinfo {author} {\bibfnamefont {M.~H.}\
  \bibnamefont {Devoret}}, \bibinfo {author} {\bibfnamefont {S.~M.}\
  \bibnamefont {Girvin}},\ and\ \bibinfo {author} {\bibfnamefont {R.~J.}\
  \bibnamefont {Schoelkopf}},\ }\bibfield  {title} {\bibinfo {title}
  {Qubit-photon interactions in a cavity: {{Measurement-induced}} dephasing and
  number splitting},\ }\href {https://doi.org/10.1103/PhysRevA.74.042318}
  {\bibfield  {journal} {\bibinfo  {journal} {Physical Review A}\ }\textbf
  {\bibinfo {volume} {74}},\ \bibinfo {pages} {042318} (\bibinfo {year}
  {2006})}\BibitemShut {NoStop}%
\bibitem [{\citenamefont {Acharya}\ \emph {et~al.}(2023)\citenamefont
  {Acharya}, \citenamefont {Aleiner}, \citenamefont {Allen}, \citenamefont
  {Andersen}, \citenamefont {Ansmann}, \citenamefont {Arute}, \citenamefont
  {Arya}, \citenamefont {Asfaw}, \citenamefont {Atalaya}, \citenamefont
  {Babbush} \emph {et~al.}}]{acharya_suppressing_2023}%
  \BibitemOpen
  \bibfield  {author} {\bibinfo {author} {\bibfnamefont {R.}~\bibnamefont
  {Acharya}}, \bibinfo {author} {\bibfnamefont {I.}~\bibnamefont {Aleiner}},
  \bibinfo {author} {\bibfnamefont {R.}~\bibnamefont {Allen}}, \bibinfo
  {author} {\bibfnamefont {T.~I.}\ \bibnamefont {Andersen}}, \bibinfo {author}
  {\bibfnamefont {M.}~\bibnamefont {Ansmann}}, \bibinfo {author} {\bibfnamefont
  {F.}~\bibnamefont {Arute}}, \bibinfo {author} {\bibfnamefont
  {K.}~\bibnamefont {Arya}}, \bibinfo {author} {\bibfnamefont {A.}~\bibnamefont
  {Asfaw}}, \bibinfo {author} {\bibfnamefont {J.}~\bibnamefont {Atalaya}},
  \bibinfo {author} {\bibfnamefont {R.}~\bibnamefont {Babbush}}, \emph
  {et~al.},\ }\bibfield  {title} {\bibinfo {title} {Suppressing quantum errors
  by scaling a surface code logical qubit},\ }\href
  {https://doi.org/10.1038/s41586-022-05434-1} {\bibfield  {journal} {\bibinfo
  {journal} {Nature}\ }\textbf {\bibinfo {volume} {614}},\ \bibinfo {pages}
  {676} (\bibinfo {year} {2023})}\BibitemShut {NoStop}%
\bibitem [{\citenamefont {Krinner}\ \emph {et~al.}(2022)\citenamefont
  {Krinner}, \citenamefont {Lacroix}, \citenamefont {Remm}, \citenamefont
  {Di~Paolo}, \citenamefont {Genois}, \citenamefont {Leroux}, \citenamefont
  {Hellings}, \citenamefont {Lazar}, \citenamefont {Swiadek}, \citenamefont
  {Herrmann}, \citenamefont {Norris}, \citenamefont {Andersen}, \citenamefont
  {M{\"u}ller}, \citenamefont {Blais}, \citenamefont {Eichler},\ and\
  \citenamefont {Wallraff}}]{krinner_realizing_2022}%
  \BibitemOpen
  \bibfield  {author} {\bibinfo {author} {\bibfnamefont {S.}~\bibnamefont
  {Krinner}}, \bibinfo {author} {\bibfnamefont {N.}~\bibnamefont {Lacroix}},
  \bibinfo {author} {\bibfnamefont {A.}~\bibnamefont {Remm}}, \bibinfo {author}
  {\bibfnamefont {A.}~\bibnamefont {Di~Paolo}}, \bibinfo {author}
  {\bibfnamefont {E.}~\bibnamefont {Genois}}, \bibinfo {author} {\bibfnamefont
  {C.}~\bibnamefont {Leroux}}, \bibinfo {author} {\bibfnamefont
  {C.}~\bibnamefont {Hellings}}, \bibinfo {author} {\bibfnamefont
  {S.}~\bibnamefont {Lazar}}, \bibinfo {author} {\bibfnamefont
  {F.}~\bibnamefont {Swiadek}}, \bibinfo {author} {\bibfnamefont
  {J.}~\bibnamefont {Herrmann}}, \bibinfo {author} {\bibfnamefont {G.~J.}\
  \bibnamefont {Norris}}, \bibinfo {author} {\bibfnamefont {C.~K.}\
  \bibnamefont {Andersen}}, \bibinfo {author} {\bibfnamefont {M.}~\bibnamefont
  {M{\"u}ller}}, \bibinfo {author} {\bibfnamefont {A.}~\bibnamefont {Blais}},
  \bibinfo {author} {\bibfnamefont {C.}~\bibnamefont {Eichler}},\ and\ \bibinfo
  {author} {\bibfnamefont {A.}~\bibnamefont {Wallraff}},\ }\bibfield  {title}
  {\bibinfo {title} {Realizing repeated quantum error correction in a
  distance-three surface code},\ }\href
  {https://doi.org/10.1038/s41586-022-04566-8} {\bibfield  {journal} {\bibinfo
  {journal} {Nature}\ }\textbf {\bibinfo {volume} {605}},\ \bibinfo {pages}
  {669} (\bibinfo {year} {2022})}\BibitemShut {NoStop}%
\bibitem [{\citenamefont {Chen}\ \emph {et~al.}(2021)\citenamefont {Chen},
  \citenamefont {Satzinger}, \citenamefont {Atalaya}, \citenamefont {Korotkov},
  \citenamefont {Dunsworth}, \citenamefont {Sank}, \citenamefont {Quintana},
  \citenamefont {McEwen}, \citenamefont {Barends}, \citenamefont {Klimov} \emph
  {et~al.}}]{chen_exponential_2021}%
  \BibitemOpen
  \bibfield  {author} {\bibinfo {author} {\bibfnamefont {Z.}~\bibnamefont
  {Chen}}, \bibinfo {author} {\bibfnamefont {K.~J.}\ \bibnamefont {Satzinger}},
  \bibinfo {author} {\bibfnamefont {J.}~\bibnamefont {Atalaya}}, \bibinfo
  {author} {\bibfnamefont {A.~N.}\ \bibnamefont {Korotkov}}, \bibinfo {author}
  {\bibfnamefont {A.}~\bibnamefont {Dunsworth}}, \bibinfo {author}
  {\bibfnamefont {D.}~\bibnamefont {Sank}}, \bibinfo {author} {\bibfnamefont
  {C.}~\bibnamefont {Quintana}}, \bibinfo {author} {\bibfnamefont
  {M.}~\bibnamefont {McEwen}}, \bibinfo {author} {\bibfnamefont
  {R.}~\bibnamefont {Barends}}, \bibinfo {author} {\bibfnamefont {P.~V.}\
  \bibnamefont {Klimov}}, \emph {et~al.},\ }\bibfield  {title} {\bibinfo
  {title} {Exponential suppression of bit or phase errors with cyclic error
  correction},\ }\href {https://doi.org/10.1038/s41586-021-03588-y} {\bibfield
  {journal} {\bibinfo  {journal} {Nature}\ }\textbf {\bibinfo {volume} {595}},\
  \bibinfo {pages} {383} (\bibinfo {year} {2021})}\BibitemShut {NoStop}%
\bibitem [{\citenamefont {Bengtsson}\ \emph {et~al.}(2024)\citenamefont
  {Bengtsson}, \citenamefont {Opremcak}, \citenamefont {Khezri}, \citenamefont
  {Sank}, \citenamefont {Bourassa}, \citenamefont {Satzinger}, \citenamefont
  {Hong}, \citenamefont {Erickson}, \citenamefont {Lester}, \citenamefont
  {Miao}, \citenamefont {Korotkov}, \citenamefont {Kelly}, \citenamefont
  {Chen},\ and\ \citenamefont {Klimov}}]{bengtsson_model-based_2024}%
  \BibitemOpen
  \bibfield  {author} {\bibinfo {author} {\bibfnamefont {A.}~\bibnamefont
  {Bengtsson}}, \bibinfo {author} {\bibfnamefont {A.}~\bibnamefont {Opremcak}},
  \bibinfo {author} {\bibfnamefont {M.}~\bibnamefont {Khezri}}, \bibinfo
  {author} {\bibfnamefont {D.}~\bibnamefont {Sank}}, \bibinfo {author}
  {\bibfnamefont {A.}~\bibnamefont {Bourassa}}, \bibinfo {author}
  {\bibfnamefont {K.~J.}\ \bibnamefont {Satzinger}}, \bibinfo {author}
  {\bibfnamefont {S.}~\bibnamefont {Hong}}, \bibinfo {author} {\bibfnamefont
  {C.}~\bibnamefont {Erickson}}, \bibinfo {author} {\bibfnamefont {B.~J.}\
  \bibnamefont {Lester}}, \bibinfo {author} {\bibfnamefont {K.~C.}\
  \bibnamefont {Miao}}, \bibinfo {author} {\bibfnamefont {A.~N.}\ \bibnamefont
  {Korotkov}}, \bibinfo {author} {\bibfnamefont {J.}~\bibnamefont {Kelly}},
  \bibinfo {author} {\bibfnamefont {Z.}~\bibnamefont {Chen}},\ and\ \bibinfo
  {author} {\bibfnamefont {P.~V.}\ \bibnamefont {Klimov}},\ }\bibfield  {title}
  {\bibinfo {title} {Model-{{Based Optimization}} of {{Superconducting Qubit
  Readout}}},\ }\href {https://doi.org/10.1103/PhysRevLett.132.100603}
  {\bibfield  {journal} {\bibinfo  {journal} {Physical Review Letters}\
  }\textbf {\bibinfo {volume} {132}},\ \bibinfo {pages} {100603} (\bibinfo
  {year} {2024})}\BibitemShut {NoStop}%
\bibitem [{\citenamefont {Purcell}(1946)}]{purcell_proceedings_1946}%
  \BibitemOpen
  \bibfield  {author} {\bibinfo {author} {\bibfnamefont {E.}~\bibnamefont
  {Purcell}},\ }\bibfield  {title} {\bibinfo {title} {Proceedings of the
  {{American Physical Society}}},\ }\href
  {https://doi.org/10.1103/PhysRev.69.674} {\bibfield  {journal} {\bibinfo
  {journal} {Physical Review}\ }\textbf {\bibinfo {volume} {69}},\ \bibinfo
  {pages} {674} (\bibinfo {year} {1946})}\BibitemShut {NoStop}%
\bibitem [{\citenamefont {Reed}\ \emph {et~al.}(2010)\citenamefont {Reed},
  \citenamefont {Johnson}, \citenamefont {Houck}, \citenamefont {DiCarlo},
  \citenamefont {Chow}, \citenamefont {Schuster}, \citenamefont {Frunzio},\
  and\ \citenamefont {Schoelkopf}}]{reed_fast_2010}%
  \BibitemOpen
  \bibfield  {author} {\bibinfo {author} {\bibfnamefont {M.~D.}\ \bibnamefont
  {Reed}}, \bibinfo {author} {\bibfnamefont {B.~R.}\ \bibnamefont {Johnson}},
  \bibinfo {author} {\bibfnamefont {A.~A.}\ \bibnamefont {Houck}}, \bibinfo
  {author} {\bibfnamefont {L.}~\bibnamefont {DiCarlo}}, \bibinfo {author}
  {\bibfnamefont {J.~M.}\ \bibnamefont {Chow}}, \bibinfo {author}
  {\bibfnamefont {D.~I.}\ \bibnamefont {Schuster}}, \bibinfo {author}
  {\bibfnamefont {L.}~\bibnamefont {Frunzio}},\ and\ \bibinfo {author}
  {\bibfnamefont {R.~J.}\ \bibnamefont {Schoelkopf}},\ }\bibfield  {title}
  {\bibinfo {title} {Fast reset and suppressing spontaneous emission of a
  superconducting qubit},\ }\href {https://doi.org/10.1063/1.3435463}
  {\bibfield  {journal} {\bibinfo  {journal} {Applied Physics Letters}\
  }\textbf {\bibinfo {volume} {96}},\ \bibinfo {pages} {203110} (\bibinfo
  {year} {2010})}\BibitemShut {NoStop}%
\bibitem [{\citenamefont {Jeffrey}\ \emph {et~al.}(2014)\citenamefont
  {Jeffrey}, \citenamefont {Sank}, \citenamefont {Mutus}, \citenamefont
  {White}, \citenamefont {Kelly}, \citenamefont {Barends}, \citenamefont
  {Chen}, \citenamefont {Chen}, \citenamefont {Chiaro}, \citenamefont
  {Dunsworth}, \citenamefont {Megrant}, \citenamefont {O'Malley}, \citenamefont
  {Neill}, \citenamefont {Roushan}, \citenamefont {Vainsencher}, \citenamefont
  {Wenner}, \citenamefont {Cleland},\ and\ \citenamefont
  {Martinis}}]{jeffrey_fast_2014}%
  \BibitemOpen
  \bibfield  {author} {\bibinfo {author} {\bibfnamefont {E.}~\bibnamefont
  {Jeffrey}}, \bibinfo {author} {\bibfnamefont {D.}~\bibnamefont {Sank}},
  \bibinfo {author} {\bibfnamefont {J.~Y.}\ \bibnamefont {Mutus}}, \bibinfo
  {author} {\bibfnamefont {T.~C.}\ \bibnamefont {White}}, \bibinfo {author}
  {\bibfnamefont {J.}~\bibnamefont {Kelly}}, \bibinfo {author} {\bibfnamefont
  {R.}~\bibnamefont {Barends}}, \bibinfo {author} {\bibfnamefont
  {Y.}~\bibnamefont {Chen}}, \bibinfo {author} {\bibfnamefont {Z.}~\bibnamefont
  {Chen}}, \bibinfo {author} {\bibfnamefont {B.}~\bibnamefont {Chiaro}},
  \bibinfo {author} {\bibfnamefont {A.}~\bibnamefont {Dunsworth}}, \bibinfo
  {author} {\bibfnamefont {A.}~\bibnamefont {Megrant}}, \bibinfo {author}
  {\bibfnamefont {P.~J.~J.}\ \bibnamefont {O'Malley}}, \bibinfo {author}
  {\bibfnamefont {C.}~\bibnamefont {Neill}}, \bibinfo {author} {\bibfnamefont
  {P.}~\bibnamefont {Roushan}}, \bibinfo {author} {\bibfnamefont
  {A.}~\bibnamefont {Vainsencher}}, \bibinfo {author} {\bibfnamefont
  {J.}~\bibnamefont {Wenner}}, \bibinfo {author} {\bibfnamefont {A.~N.}\
  \bibnamefont {Cleland}},\ and\ \bibinfo {author} {\bibfnamefont {J.~M.}\
  \bibnamefont {Martinis}},\ }\bibfield  {title} {\bibinfo {title} {Fast
  {{Accurate State Measurement}} with {{Superconducting Qubits}}},\ }\href
  {https://doi.org/10.1103/PhysRevLett.112.190504} {\bibfield  {journal}
  {\bibinfo  {journal} {Physical Review Letters}\ }\textbf {\bibinfo {volume}
  {112}},\ \bibinfo {pages} {190504} (\bibinfo {year} {2014})}\BibitemShut
  {NoStop}%
\bibitem [{\citenamefont {Sete}\ \emph {et~al.}(2015)\citenamefont {Sete},
  \citenamefont {Martinis},\ and\ \citenamefont
  {Korotkov}}]{sete_quantum_2015}%
  \BibitemOpen
  \bibfield  {author} {\bibinfo {author} {\bibfnamefont {E.~A.}\ \bibnamefont
  {Sete}}, \bibinfo {author} {\bibfnamefont {J.~M.}\ \bibnamefont {Martinis}},\
  and\ \bibinfo {author} {\bibfnamefont {A.~N.}\ \bibnamefont {Korotkov}},\
  }\bibfield  {title} {\bibinfo {title} {Quantum theory of a bandpass
  {{Purcell}} filter for qubit readout},\ }\href
  {https://doi.org/10.1103/PhysRevA.92.012325} {\bibfield  {journal} {\bibinfo
  {journal} {Physical Review A}\ }\textbf {\bibinfo {volume} {92}},\ \bibinfo
  {pages} {012325} (\bibinfo {year} {2015})}\BibitemShut {NoStop}%
\bibitem [{\citenamefont {Bronn}\ \emph
  {et~al.}(2015{\natexlab{a}})\citenamefont {Bronn}, \citenamefont {Liu},
  \citenamefont {Hertzberg}, \citenamefont {C{\'o}rcoles}, \citenamefont
  {Houck}, \citenamefont {Gambetta},\ and\ \citenamefont
  {Chow}}]{bronn_broadband_2015}%
  \BibitemOpen
  \bibfield  {author} {\bibinfo {author} {\bibfnamefont {N.~T.}\ \bibnamefont
  {Bronn}}, \bibinfo {author} {\bibfnamefont {Y.}~\bibnamefont {Liu}}, \bibinfo
  {author} {\bibfnamefont {J.~B.}\ \bibnamefont {Hertzberg}}, \bibinfo {author}
  {\bibfnamefont {A.~D.}\ \bibnamefont {C{\'o}rcoles}}, \bibinfo {author}
  {\bibfnamefont {A.~A.}\ \bibnamefont {Houck}}, \bibinfo {author}
  {\bibfnamefont {J.~M.}\ \bibnamefont {Gambetta}},\ and\ \bibinfo {author}
  {\bibfnamefont {J.~M.}\ \bibnamefont {Chow}},\ }\bibfield  {title} {\bibinfo
  {title} {Broadband filters for abatement of spontaneous emission in circuit
  quantum electrodynamics},\ }\href {https://doi.org/10.1063/1.4934867}
  {\bibfield  {journal} {\bibinfo  {journal} {Applied Physics Letters}\
  }\textbf {\bibinfo {volume} {107}},\ \bibinfo {pages} {172601} (\bibinfo
  {year} {2015}{\natexlab{a}})}\BibitemShut {NoStop}%
\bibitem [{\citenamefont {Cleland}\ \emph {et~al.}(2019)\citenamefont
  {Cleland}, \citenamefont {Pechal}, \citenamefont {Stas}, \citenamefont
  {Sarabalis}, \citenamefont {Wollack},\ and\ \citenamefont
  {{Safavi-Naeini}}}]{cleland_mechanical_2019}%
  \BibitemOpen
  \bibfield  {author} {\bibinfo {author} {\bibfnamefont {A.~Y.}\ \bibnamefont
  {Cleland}}, \bibinfo {author} {\bibfnamefont {M.}~\bibnamefont {Pechal}},
  \bibinfo {author} {\bibfnamefont {P.-J.~C.}\ \bibnamefont {Stas}}, \bibinfo
  {author} {\bibfnamefont {C.~J.}\ \bibnamefont {Sarabalis}}, \bibinfo {author}
  {\bibfnamefont {E.~A.}\ \bibnamefont {Wollack}},\ and\ \bibinfo {author}
  {\bibfnamefont {A.~H.}\ \bibnamefont {{Safavi-Naeini}}},\ }\bibfield  {title}
  {\bibinfo {title} {Mechanical {{Purcell}} filters for microwave quantum
  machines},\ }\href {https://doi.org/10.1063/1.5111151} {\bibfield  {journal}
  {\bibinfo  {journal} {Applied Physics Letters}\ }\textbf {\bibinfo {volume}
  {115}},\ \bibinfo {pages} {263504} (\bibinfo {year} {2019})}\BibitemShut
  {NoStop}%
\bibitem [{\citenamefont {Yan}\ \emph {et~al.}(2023)\citenamefont {Yan},
  \citenamefont {Wu}, \citenamefont {Lingenfelter}, \citenamefont {Joshi},
  \citenamefont {Andersson}, \citenamefont {Conner}, \citenamefont {Chou},
  \citenamefont {Grebel}, \citenamefont {Miller}, \citenamefont {Povey},
  \citenamefont {Qiao}, \citenamefont {Clerk},\ and\ \citenamefont
  {Cleland}}]{yan_broadband_2023}%
  \BibitemOpen
  \bibfield  {author} {\bibinfo {author} {\bibfnamefont {H.}~\bibnamefont
  {Yan}}, \bibinfo {author} {\bibfnamefont {X.}~\bibnamefont {Wu}}, \bibinfo
  {author} {\bibfnamefont {A.}~\bibnamefont {Lingenfelter}}, \bibinfo {author}
  {\bibfnamefont {Y.~J.}\ \bibnamefont {Joshi}}, \bibinfo {author}
  {\bibfnamefont {G.}~\bibnamefont {Andersson}}, \bibinfo {author}
  {\bibfnamefont {C.~R.}\ \bibnamefont {Conner}}, \bibinfo {author}
  {\bibfnamefont {M.-H.}\ \bibnamefont {Chou}}, \bibinfo {author}
  {\bibfnamefont {J.}~\bibnamefont {Grebel}}, \bibinfo {author} {\bibfnamefont
  {J.~M.}\ \bibnamefont {Miller}}, \bibinfo {author} {\bibfnamefont {R.~G.}\
  \bibnamefont {Povey}}, \bibinfo {author} {\bibfnamefont {H.}~\bibnamefont
  {Qiao}}, \bibinfo {author} {\bibfnamefont {A.~A.}\ \bibnamefont {Clerk}},\
  and\ \bibinfo {author} {\bibfnamefont {A.~N.}\ \bibnamefont {Cleland}},\
  }\bibfield  {title} {\bibinfo {title} {Broadband bandpass {{Purcell}} filter
  for circuit quantum electrodynamics},\ }\href
  {https://doi.org/10.1063/5.0161893} {\bibfield  {journal} {\bibinfo
  {journal} {Applied Physics Letters}\ }\textbf {\bibinfo {volume} {123}},\
  \bibinfo {pages} {134001} (\bibinfo {year} {2023})}\BibitemShut {NoStop}%
\bibitem [{\citenamefont {Park}\ \emph {et~al.}(2024)\citenamefont {Park},
  \citenamefont {Choi}, \citenamefont {Kim}, \citenamefont {Jo}, \citenamefont
  {Lee}, \citenamefont {Kim}, \citenamefont {Park}, \citenamefont {Lee},\ and\
  \citenamefont {Hahn}}]{park_characterization_2024}%
  \BibitemOpen
  \bibfield  {author} {\bibinfo {author} {\bibfnamefont {S.~H.}\ \bibnamefont
  {Park}}, \bibinfo {author} {\bibfnamefont {G.}~\bibnamefont {Choi}}, \bibinfo
  {author} {\bibfnamefont {G.}~\bibnamefont {Kim}}, \bibinfo {author}
  {\bibfnamefont {J.}~\bibnamefont {Jo}}, \bibinfo {author} {\bibfnamefont
  {B.}~\bibnamefont {Lee}}, \bibinfo {author} {\bibfnamefont {G.}~\bibnamefont
  {Kim}}, \bibinfo {author} {\bibfnamefont {K.}~\bibnamefont {Park}}, \bibinfo
  {author} {\bibfnamefont {Y.-H.}\ \bibnamefont {Lee}},\ and\ \bibinfo {author}
  {\bibfnamefont {S.}~\bibnamefont {Hahn}},\ }\bibfield  {title} {\bibinfo
  {title} {Characterization of broadband {{Purcell}} filters with compact
  footprint for fast multiplexed superconducting qubit readout},\ }\href
  {https://doi.org/10.1063/5.0182642} {\bibfield  {journal} {\bibinfo
  {journal} {Applied Physics Letters}\ }\textbf {\bibinfo {volume} {124}},\
  \bibinfo {pages} {044003} (\bibinfo {year} {2024})}\BibitemShut {NoStop}%
\bibitem [{\citenamefont {Govia}\ and\ \citenamefont
  {Clerk}(2017)}]{govia_enhanced_2017}%
  \BibitemOpen
  \bibfield  {author} {\bibinfo {author} {\bibfnamefont {L.~C.~G.}\
  \bibnamefont {Govia}}\ and\ \bibinfo {author} {\bibfnamefont {A.~A.}\
  \bibnamefont {Clerk}},\ }\bibfield  {title} {\bibinfo {title} {Enhanced qubit
  readout using locally generated squeezing and inbuilt {{Purcell-decay}}
  suppression},\ }\href {https://doi.org/10.1088/1367-2630/aa5f7b} {\bibfield
  {journal} {\bibinfo  {journal} {New Journal of Physics}\ }\textbf {\bibinfo
  {volume} {19}},\ \bibinfo {pages} {023044} (\bibinfo {year}
  {2017})}\BibitemShut {NoStop}%
\bibitem [{\citenamefont {Bronn}\ \emph
  {et~al.}(2015{\natexlab{b}})\citenamefont {Bronn}, \citenamefont {Magesan},
  \citenamefont {Masluk}, \citenamefont {Chow}, \citenamefont {Gambetta},\ and\
  \citenamefont {Steffen}}]{bronn_reducing_2015}%
  \BibitemOpen
  \bibfield  {author} {\bibinfo {author} {\bibfnamefont {N.~T.}\ \bibnamefont
  {Bronn}}, \bibinfo {author} {\bibfnamefont {E.}~\bibnamefont {Magesan}},
  \bibinfo {author} {\bibfnamefont {N.~A.}\ \bibnamefont {Masluk}}, \bibinfo
  {author} {\bibfnamefont {J.~M.}\ \bibnamefont {Chow}}, \bibinfo {author}
  {\bibfnamefont {J.~M.}\ \bibnamefont {Gambetta}},\ and\ \bibinfo {author}
  {\bibfnamefont {M.}~\bibnamefont {Steffen}},\ }\bibfield  {title} {\bibinfo
  {title} {Reducing {{Spontaneous Emission}} in {{Circuit Quantum
  Electrodynamics}} by a {{Combined Readout}}/{{Filter Technique}}},\ }\href
  {https://doi.org/10.1109/TASC.2015.2456109} {\bibfield  {journal} {\bibinfo
  {journal} {IEEE Transactions on Applied Superconductivity}\ }\textbf
  {\bibinfo {volume} {25}},\ \bibinfo {pages} {1} (\bibinfo {year}
  {2015}{\natexlab{b}})}\BibitemShut {NoStop}%
\bibitem [{\citenamefont {Sunada}\ \emph {et~al.}(2022)\citenamefont {Sunada},
  \citenamefont {Kono}, \citenamefont {Ilves}, \citenamefont {Tamate},
  \citenamefont {Sugiyama}, \citenamefont {Tabuchi},\ and\ \citenamefont
  {Nakamura}}]{sunada_fast_2022}%
  \BibitemOpen
  \bibfield  {author} {\bibinfo {author} {\bibfnamefont {Y.}~\bibnamefont
  {Sunada}}, \bibinfo {author} {\bibfnamefont {S.}~\bibnamefont {Kono}},
  \bibinfo {author} {\bibfnamefont {J.}~\bibnamefont {Ilves}}, \bibinfo
  {author} {\bibfnamefont {S.}~\bibnamefont {Tamate}}, \bibinfo {author}
  {\bibfnamefont {T.}~\bibnamefont {Sugiyama}}, \bibinfo {author}
  {\bibfnamefont {Y.}~\bibnamefont {Tabuchi}},\ and\ \bibinfo {author}
  {\bibfnamefont {Y.}~\bibnamefont {Nakamura}},\ }\bibfield  {title} {\bibinfo
  {title} {Fast {{Readout}} and {{Reset}} of a {{Superconducting Qubit
  Coupled}} to a {{Resonator}} with an {{Intrinsic Purcell Filter}}},\ }\href
  {https://doi.org/10.1103/PhysRevApplied.17.044016} {\bibfield  {journal}
  {\bibinfo  {journal} {Physical Review Applied}\ }\textbf {\bibinfo {volume}
  {17}},\ \bibinfo {pages} {044016} (\bibinfo {year} {2022})}\BibitemShut
  {NoStop}%
\bibitem [{\citenamefont {Yen}\ \emph {et~al.}(2024)\citenamefont {Yen},
  \citenamefont {Ye}, \citenamefont {Peng}, \citenamefont {Wang}, \citenamefont
  {Cunningham}, \citenamefont {Gingras}, \citenamefont {Niedzielski},
  \citenamefont {Stickler}, \citenamefont {Serniak}, \citenamefont {Schwartz},\
  and\ \citenamefont {O'Brien}}]{yen_directional_2024}%
  \BibitemOpen
  \bibfield  {author} {\bibinfo {author} {\bibfnamefont {A.}~\bibnamefont
  {Yen}}, \bibinfo {author} {\bibfnamefont {Y.}~\bibnamefont {Ye}}, \bibinfo
  {author} {\bibfnamefont {K.}~\bibnamefont {Peng}}, \bibinfo {author}
  {\bibfnamefont {J.}~\bibnamefont {Wang}}, \bibinfo {author} {\bibfnamefont
  {G.}~\bibnamefont {Cunningham}}, \bibinfo {author} {\bibfnamefont
  {M.}~\bibnamefont {Gingras}}, \bibinfo {author} {\bibfnamefont {B.~M.}\
  \bibnamefont {Niedzielski}}, \bibinfo {author} {\bibfnamefont
  {H.}~\bibnamefont {Stickler}}, \bibinfo {author} {\bibfnamefont
  {K.}~\bibnamefont {Serniak}}, \bibinfo {author} {\bibfnamefont {M.~E.}\
  \bibnamefont {Schwartz}},\ and\ \bibinfo {author} {\bibfnamefont {K.~P.}\
  \bibnamefont {O'Brien}},\ }\bibfield  {title} {\bibinfo {title} {Directional
  emission of a readout resonator for qubit measurement},\ }\href
  {https://doi.org/10.1103/PhysRevApplied.22.034035} {\bibfield  {journal}
  {\bibinfo  {journal} {Physical Review Applied}\ }\textbf {\bibinfo {volume}
  {22}},\ \bibinfo {pages} {034035} (\bibinfo {year} {2024})}\BibitemShut
  {NoStop}%
\bibitem [{\citenamefont {Koch}\ \emph {et~al.}(2007)\citenamefont {Koch},
  \citenamefont {Yu}, \citenamefont {Gambetta}, \citenamefont {Houck},
  \citenamefont {Schuster}, \citenamefont {Majer}, \citenamefont {Blais},
  \citenamefont {Devoret}, \citenamefont {Girvin},\ and\ \citenamefont
  {Schoelkopf}}]{koch_charge-insensitive_2007}%
  \BibitemOpen
  \bibfield  {author} {\bibinfo {author} {\bibfnamefont {J.}~\bibnamefont
  {Koch}}, \bibinfo {author} {\bibfnamefont {T.~M.}\ \bibnamefont {Yu}},
  \bibinfo {author} {\bibfnamefont {J.}~\bibnamefont {Gambetta}}, \bibinfo
  {author} {\bibfnamefont {A.~A.}\ \bibnamefont {Houck}}, \bibinfo {author}
  {\bibfnamefont {D.~I.}\ \bibnamefont {Schuster}}, \bibinfo {author}
  {\bibfnamefont {J.}~\bibnamefont {Majer}}, \bibinfo {author} {\bibfnamefont
  {A.}~\bibnamefont {Blais}}, \bibinfo {author} {\bibfnamefont {M.~H.}\
  \bibnamefont {Devoret}}, \bibinfo {author} {\bibfnamefont {S.~M.}\
  \bibnamefont {Girvin}},\ and\ \bibinfo {author} {\bibfnamefont {R.~J.}\
  \bibnamefont {Schoelkopf}},\ }\bibfield  {title} {\bibinfo {title}
  {Charge-insensitive qubit design derived from the {{Cooper}} pair box},\
  }\href {https://doi.org/10.1103/PhysRevA.76.042319} {\bibfield  {journal}
  {\bibinfo  {journal} {Physical Review A}\ }\textbf {\bibinfo {volume} {76}},\
  \bibinfo {pages} {042319} (\bibinfo {year} {2007})}\BibitemShut {NoStop}%
\bibitem [{\citenamefont {Kockum}\ \emph {et~al.}(2018)\citenamefont {Kockum},
  \citenamefont {Johansson},\ and\ \citenamefont
  {Nori}}]{kockum_decoherence-free_2018}%
  \BibitemOpen
  \bibfield  {author} {\bibinfo {author} {\bibfnamefont {A.~F.}\ \bibnamefont
  {Kockum}}, \bibinfo {author} {\bibfnamefont {G.}~\bibnamefont {Johansson}},\
  and\ \bibinfo {author} {\bibfnamefont {F.}~\bibnamefont {Nori}},\ }\bibfield
  {title} {\bibinfo {title} {Decoherence-{{Free Interaction}} between {{Giant
  Atoms}} in {{Waveguide Quantum Electrodynamics}}},\ }\href
  {https://doi.org/10.1103/PhysRevLett.120.140404} {\bibfield  {journal}
  {\bibinfo  {journal} {Physical Review Letters}\ }\textbf {\bibinfo {volume}
  {120}},\ \bibinfo {pages} {140404} (\bibinfo {year} {2018})}\BibitemShut
  {NoStop}%
\bibitem [{\citenamefont {Kannan}\ \emph {et~al.}(2020)\citenamefont {Kannan},
  \citenamefont {Ruckriegel}, \citenamefont {Campbell}, \citenamefont
  {Frisk~Kockum}, \citenamefont {Braum{\"u}ller}, \citenamefont {Kim},
  \citenamefont {Kjaergaard}, \citenamefont {Krantz}, \citenamefont {Melville},
  \citenamefont {Niedzielski}, \citenamefont {Veps{\"a}l{\"a}inen},
  \citenamefont {Winik}, \citenamefont {Yoder}, \citenamefont {Nori},
  \citenamefont {Orlando}, \citenamefont {Gustavsson},\ and\ \citenamefont
  {Oliver}}]{kannan_waveguide_2020}%
  \BibitemOpen
  \bibfield  {author} {\bibinfo {author} {\bibfnamefont {B.}~\bibnamefont
  {Kannan}}, \bibinfo {author} {\bibfnamefont {M.~J.}\ \bibnamefont
  {Ruckriegel}}, \bibinfo {author} {\bibfnamefont {D.~L.}\ \bibnamefont
  {Campbell}}, \bibinfo {author} {\bibfnamefont {A.}~\bibnamefont
  {Frisk~Kockum}}, \bibinfo {author} {\bibfnamefont {J.}~\bibnamefont
  {Braum{\"u}ller}}, \bibinfo {author} {\bibfnamefont {D.~K.}\ \bibnamefont
  {Kim}}, \bibinfo {author} {\bibfnamefont {M.}~\bibnamefont {Kjaergaard}},
  \bibinfo {author} {\bibfnamefont {P.}~\bibnamefont {Krantz}}, \bibinfo
  {author} {\bibfnamefont {A.}~\bibnamefont {Melville}}, \bibinfo {author}
  {\bibfnamefont {B.~M.}\ \bibnamefont {Niedzielski}}, \bibinfo {author}
  {\bibfnamefont {A.}~\bibnamefont {Veps{\"a}l{\"a}inen}}, \bibinfo {author}
  {\bibfnamefont {R.}~\bibnamefont {Winik}}, \bibinfo {author} {\bibfnamefont
  {J.~L.}\ \bibnamefont {Yoder}}, \bibinfo {author} {\bibfnamefont
  {F.}~\bibnamefont {Nori}}, \bibinfo {author} {\bibfnamefont {T.~P.}\
  \bibnamefont {Orlando}}, \bibinfo {author} {\bibfnamefont {S.}~\bibnamefont
  {Gustavsson}},\ and\ \bibinfo {author} {\bibfnamefont {W.~D.}\ \bibnamefont
  {Oliver}},\ }\bibfield  {title} {\bibinfo {title} {Waveguide quantum
  electrodynamics with superconducting artificial giant atoms},\ }\href
  {https://doi.org/10.1038/s41586-020-2529-9} {\bibfield  {journal} {\bibinfo
  {journal} {Nature}\ }\textbf {\bibinfo {volume} {583}},\ \bibinfo {pages}
  {775} (\bibinfo {year} {2020})}\BibitemShut {NoStop}%
\bibitem [{\citenamefont {Esteve}\ \emph {et~al.}(1986)\citenamefont {Esteve},
  \citenamefont {Devoret},\ and\ \citenamefont
  {Martinis}}]{esteve_effect_1986}%
  \BibitemOpen
  \bibfield  {author} {\bibinfo {author} {\bibfnamefont {D.}~\bibnamefont
  {Esteve}}, \bibinfo {author} {\bibfnamefont {M.~H.}\ \bibnamefont
  {Devoret}},\ and\ \bibinfo {author} {\bibfnamefont {J.~M.}\ \bibnamefont
  {Martinis}},\ }\bibfield  {title} {\bibinfo {title} {Effect of an arbitrary
  dissipative circuit on the quantum energy levels and tunneling of a
  {{Josephson}} junction},\ }\href {https://doi.org/10.1103/PhysRevB.34.158}
  {\bibfield  {journal} {\bibinfo  {journal} {Physical Review B}\ }\textbf
  {\bibinfo {volume} {34}},\ \bibinfo {pages} {158} (\bibinfo {year}
  {1986})}\BibitemShut {NoStop}%
\bibitem [{\citenamefont {Neeley}\ \emph {et~al.}(2008)\citenamefont {Neeley},
  \citenamefont {Ansmann}, \citenamefont {Bialczak}, \citenamefont {Hofheinz},
  \citenamefont {Katz}, \citenamefont {Lucero}, \citenamefont {O'Connell},
  \citenamefont {Wang}, \citenamefont {Cleland},\ and\ \citenamefont
  {Martinis}}]{neeley_transformed_2008}%
  \BibitemOpen
  \bibfield  {author} {\bibinfo {author} {\bibfnamefont {M.}~\bibnamefont
  {Neeley}}, \bibinfo {author} {\bibfnamefont {M.}~\bibnamefont {Ansmann}},
  \bibinfo {author} {\bibfnamefont {R.~C.}\ \bibnamefont {Bialczak}}, \bibinfo
  {author} {\bibfnamefont {M.}~\bibnamefont {Hofheinz}}, \bibinfo {author}
  {\bibfnamefont {N.}~\bibnamefont {Katz}}, \bibinfo {author} {\bibfnamefont
  {E.}~\bibnamefont {Lucero}}, \bibinfo {author} {\bibfnamefont
  {A.}~\bibnamefont {O'Connell}}, \bibinfo {author} {\bibfnamefont
  {H.}~\bibnamefont {Wang}}, \bibinfo {author} {\bibfnamefont {A.~N.}\
  \bibnamefont {Cleland}},\ and\ \bibinfo {author} {\bibfnamefont {J.~M.}\
  \bibnamefont {Martinis}},\ }\bibfield  {title} {\bibinfo {title} {Transformed
  dissipation in superconducting quantum circuits},\ }\href
  {https://doi.org/10.1103/PhysRevB.77.180508} {\bibfield  {journal} {\bibinfo
  {journal} {Physical Review B}\ }\textbf {\bibinfo {volume} {77}},\ \bibinfo
  {pages} {180508(R)} (\bibinfo {year} {2008})}\BibitemShut {NoStop}%
\bibitem [{\citenamefont {Walter}\ \emph {et~al.}(2017)\citenamefont {Walter},
  \citenamefont {Kurpiers}, \citenamefont {Gasparinetti}, \citenamefont
  {Magnard}, \citenamefont {Poto{\v c}nik}, \citenamefont {Salath{\'e}},
  \citenamefont {Pechal}, \citenamefont {Mondal}, \citenamefont {Oppliger},
  \citenamefont {Eichler},\ and\ \citenamefont {Wallraff}}]{walter_rapid_2017}%
  \BibitemOpen
  \bibfield  {author} {\bibinfo {author} {\bibfnamefont {T.}~\bibnamefont
  {Walter}}, \bibinfo {author} {\bibfnamefont {P.}~\bibnamefont {Kurpiers}},
  \bibinfo {author} {\bibfnamefont {S.}~\bibnamefont {Gasparinetti}}, \bibinfo
  {author} {\bibfnamefont {P.}~\bibnamefont {Magnard}}, \bibinfo {author}
  {\bibfnamefont {A.}~\bibnamefont {Poto{\v c}nik}}, \bibinfo {author}
  {\bibfnamefont {Y.}~\bibnamefont {Salath{\'e}}}, \bibinfo {author}
  {\bibfnamefont {M.}~\bibnamefont {Pechal}}, \bibinfo {author} {\bibfnamefont
  {M.}~\bibnamefont {Mondal}}, \bibinfo {author} {\bibfnamefont
  {M.}~\bibnamefont {Oppliger}}, \bibinfo {author} {\bibfnamefont
  {C.}~\bibnamefont {Eichler}},\ and\ \bibinfo {author} {\bibfnamefont
  {A.}~\bibnamefont {Wallraff}},\ }\bibfield  {title} {\bibinfo {title} {Rapid
  {{High-Fidelity Single-Shot Dispersive Readout}} of {{Superconducting
  Qubits}}},\ }\href {https://doi.org/10.1103/PhysRevApplied.7.054020}
  {\bibfield  {journal} {\bibinfo  {journal} {Physical Review Applied}\
  }\textbf {\bibinfo {volume} {7}},\ \bibinfo {pages} {054020} (\bibinfo {year}
  {2017})}\BibitemShut {NoStop}%
\bibitem [{\citenamefont {Steffen}\ \emph {et~al.}(2010)\citenamefont
  {Steffen}, \citenamefont {Kumar}, \citenamefont {DiVincenzo}, \citenamefont
  {Rozen}, \citenamefont {Keefe}, \citenamefont {Rothwell},\ and\ \citenamefont
  {Ketchen}}]{steffen_high-coherence_2010}%
  \BibitemOpen
  \bibfield  {author} {\bibinfo {author} {\bibfnamefont {M.}~\bibnamefont
  {Steffen}}, \bibinfo {author} {\bibfnamefont {S.}~\bibnamefont {Kumar}},
  \bibinfo {author} {\bibfnamefont {D.~P.}\ \bibnamefont {DiVincenzo}},
  \bibinfo {author} {\bibfnamefont {J.~R.}\ \bibnamefont {Rozen}}, \bibinfo
  {author} {\bibfnamefont {G.~A.}\ \bibnamefont {Keefe}}, \bibinfo {author}
  {\bibfnamefont {M.~B.}\ \bibnamefont {Rothwell}},\ and\ \bibinfo {author}
  {\bibfnamefont {M.~B.}\ \bibnamefont {Ketchen}},\ }\bibfield  {title}
  {\bibinfo {title} {High-{{Coherence Hybrid Superconducting Qubit}}},\ }\href
  {https://doi.org/10.1103/PhysRevLett.105.100502} {\bibfield  {journal}
  {\bibinfo  {journal} {Physical Review Letters}\ }\textbf {\bibinfo {volume}
  {105}},\ \bibinfo {pages} {100502} (\bibinfo {year} {2010})}\BibitemShut
  {NoStop}%
\bibitem [{\citenamefont {Houck}\ \emph {et~al.}(2008)\citenamefont {Houck},
  \citenamefont {Schreier}, \citenamefont {Johnson}, \citenamefont {Chow},
  \citenamefont {Koch}, \citenamefont {Gambetta}, \citenamefont {Schuster},
  \citenamefont {Frunzio}, \citenamefont {Devoret}, \citenamefont {Girvin},\
  and\ \citenamefont {Schoelkopf}}]{houck_controlling_2008}%
  \BibitemOpen
  \bibfield  {author} {\bibinfo {author} {\bibfnamefont {A.~A.}\ \bibnamefont
  {Houck}}, \bibinfo {author} {\bibfnamefont {J.~A.}\ \bibnamefont {Schreier}},
  \bibinfo {author} {\bibfnamefont {B.~R.}\ \bibnamefont {Johnson}}, \bibinfo
  {author} {\bibfnamefont {J.~M.}\ \bibnamefont {Chow}}, \bibinfo {author}
  {\bibfnamefont {J.}~\bibnamefont {Koch}}, \bibinfo {author} {\bibfnamefont
  {J.~M.}\ \bibnamefont {Gambetta}}, \bibinfo {author} {\bibfnamefont {D.~I.}\
  \bibnamefont {Schuster}}, \bibinfo {author} {\bibfnamefont {L.}~\bibnamefont
  {Frunzio}}, \bibinfo {author} {\bibfnamefont {M.~H.}\ \bibnamefont
  {Devoret}}, \bibinfo {author} {\bibfnamefont {S.~M.}\ \bibnamefont
  {Girvin}},\ and\ \bibinfo {author} {\bibfnamefont {R.~J.}\ \bibnamefont
  {Schoelkopf}},\ }\bibfield  {title} {\bibinfo {title} {Controlling the
  {{Spontaneous Emission}} of a {{Superconducting Transmon Qubit}}},\ }\href
  {https://doi.org/10.1103/PhysRevLett.101.080502} {\bibfield  {journal}
  {\bibinfo  {journal} {Physical Review Letters}\ }\textbf {\bibinfo {volume}
  {101}},\ \bibinfo {pages} {080502} (\bibinfo {year} {2008})}\BibitemShut
  {NoStop}%
\bibitem [{\citenamefont {Pozar}(2012)}]{pozar_microwave_2012}%
  \BibitemOpen
  \bibfield  {author} {\bibinfo {author} {\bibfnamefont {D.~M.}\ \bibnamefont
  {Pozar}},\ }\href@noop {} {\emph {\bibinfo {title} {Microwave
  Engineering}}},\ \bibinfo {edition} {4th}\ ed.\ (\bibinfo  {publisher}
  {Wiley},\ \bibinfo {address} {Hoboken, NJ},\ \bibinfo {year}
  {2012})\BibitemShut {NoStop}%
\bibitem [{\citenamefont {Heinsoo}\ \emph {et~al.}(2018)\citenamefont
  {Heinsoo}, \citenamefont {Andersen}, \citenamefont {Remm}, \citenamefont
  {Krinner}, \citenamefont {Walter}, \citenamefont {Salath{\'e}}, \citenamefont
  {Gasparinetti}, \citenamefont {Besse}, \citenamefont {Poto{\v c}nik},
  \citenamefont {Wallraff},\ and\ \citenamefont
  {Eichler}}]{heinsoo_rapid_2018}%
  \BibitemOpen
  \bibfield  {author} {\bibinfo {author} {\bibfnamefont {J.}~\bibnamefont
  {Heinsoo}}, \bibinfo {author} {\bibfnamefont {C.~K.}\ \bibnamefont
  {Andersen}}, \bibinfo {author} {\bibfnamefont {A.}~\bibnamefont {Remm}},
  \bibinfo {author} {\bibfnamefont {S.}~\bibnamefont {Krinner}}, \bibinfo
  {author} {\bibfnamefont {T.}~\bibnamefont {Walter}}, \bibinfo {author}
  {\bibfnamefont {Y.}~\bibnamefont {Salath{\'e}}}, \bibinfo {author}
  {\bibfnamefont {S.}~\bibnamefont {Gasparinetti}}, \bibinfo {author}
  {\bibfnamefont {J.-C.}\ \bibnamefont {Besse}}, \bibinfo {author}
  {\bibfnamefont {A.}~\bibnamefont {Poto{\v c}nik}}, \bibinfo {author}
  {\bibfnamefont {A.}~\bibnamefont {Wallraff}},\ and\ \bibinfo {author}
  {\bibfnamefont {C.}~\bibnamefont {Eichler}},\ }\bibfield  {title} {\bibinfo
  {title} {Rapid {{High-fidelity Multiplexed Readout}} of {{Superconducting
  Qubits}}},\ }\href {https://doi.org/10.1103/PhysRevApplied.10.034040}
  {\bibfield  {journal} {\bibinfo  {journal} {Physical Review Applied}\
  }\textbf {\bibinfo {volume} {10}},\ \bibinfo {pages} {034040} (\bibinfo
  {year} {2018})}\BibitemShut {NoStop}%
\bibitem [{\citenamefont {Probst}\ \emph {et~al.}(2015)\citenamefont {Probst},
  \citenamefont {Song}, \citenamefont {Bushev}, \citenamefont {Ustinov},\ and\
  \citenamefont {Weides}}]{probst_efficient_2015}%
  \BibitemOpen
  \bibfield  {author} {\bibinfo {author} {\bibfnamefont {S.}~\bibnamefont
  {Probst}}, \bibinfo {author} {\bibfnamefont {F.~B.}\ \bibnamefont {Song}},
  \bibinfo {author} {\bibfnamefont {P.~A.}\ \bibnamefont {Bushev}}, \bibinfo
  {author} {\bibfnamefont {A.~V.}\ \bibnamefont {Ustinov}},\ and\ \bibinfo
  {author} {\bibfnamefont {M.}~\bibnamefont {Weides}},\ }\bibfield  {title}
  {\bibinfo {title} {Efficient and robust analysis of complex scattering data
  under noise in microwave resonators},\ }\href
  {https://doi.org/10.1063/1.4907935} {\bibfield  {journal} {\bibinfo
  {journal} {Review of Scientific Instruments}\ }\textbf {\bibinfo {volume}
  {86}},\ \bibinfo {pages} {024706} (\bibinfo {year} {2015})}\BibitemShut
  {NoStop}%
\bibitem [{\citenamefont {Stefanazzi}\ \emph {et~al.}(2022)\citenamefont
  {Stefanazzi}, \citenamefont {Treptow}, \citenamefont {Wilcer}, \citenamefont
  {Stoughton}, \citenamefont {Bradford}, \citenamefont {Uemura}, \citenamefont
  {Zorzetti}, \citenamefont {Montella}, \citenamefont {Cancelo}, \citenamefont
  {Sussman}, \citenamefont {Houck}, \citenamefont {Saxena}, \citenamefont
  {Arnaldi}, \citenamefont {Agrawal}, \citenamefont {Zhang}, \citenamefont
  {Ding},\ and\ \citenamefont {Schuster}}]{stefanazzi_qick_2022}%
  \BibitemOpen
  \bibfield  {author} {\bibinfo {author} {\bibfnamefont {L.}~\bibnamefont
  {Stefanazzi}}, \bibinfo {author} {\bibfnamefont {K.}~\bibnamefont {Treptow}},
  \bibinfo {author} {\bibfnamefont {N.}~\bibnamefont {Wilcer}}, \bibinfo
  {author} {\bibfnamefont {C.}~\bibnamefont {Stoughton}}, \bibinfo {author}
  {\bibfnamefont {C.}~\bibnamefont {Bradford}}, \bibinfo {author}
  {\bibfnamefont {S.}~\bibnamefont {Uemura}}, \bibinfo {author} {\bibfnamefont
  {S.}~\bibnamefont {Zorzetti}}, \bibinfo {author} {\bibfnamefont
  {S.}~\bibnamefont {Montella}}, \bibinfo {author} {\bibfnamefont
  {G.}~\bibnamefont {Cancelo}}, \bibinfo {author} {\bibfnamefont
  {S.}~\bibnamefont {Sussman}}, \bibinfo {author} {\bibfnamefont
  {A.}~\bibnamefont {Houck}}, \bibinfo {author} {\bibfnamefont
  {S.}~\bibnamefont {Saxena}}, \bibinfo {author} {\bibfnamefont
  {H.}~\bibnamefont {Arnaldi}}, \bibinfo {author} {\bibfnamefont
  {A.}~\bibnamefont {Agrawal}}, \bibinfo {author} {\bibfnamefont
  {H.}~\bibnamefont {Zhang}}, \bibinfo {author} {\bibfnamefont
  {C.}~\bibnamefont {Ding}},\ and\ \bibinfo {author} {\bibfnamefont {D.~I.}\
  \bibnamefont {Schuster}},\ }\bibfield  {title} {\bibinfo {title} {The
  {{QICK}} ({{Quantum Instrumentation Control Kit}}): {{Readout}} and control
  for qubits and detectors},\ }\href {https://doi.org/10.1063/5.0076249}
  {\bibfield  {journal} {\bibinfo  {journal} {Review of Scientific
  Instruments}\ }\textbf {\bibinfo {volume} {93}},\ \bibinfo {pages} {044709}
  (\bibinfo {year} {2022})}\BibitemShut {NoStop}%
\bibitem [{\citenamefont {Blais}\ \emph {et~al.}(2021)\citenamefont {Blais},
  \citenamefont {Grimsmo}, \citenamefont {Girvin},\ and\ \citenamefont
  {Wallraff}}]{blais_circuit_2021}%
  \BibitemOpen
  \bibfield  {author} {\bibinfo {author} {\bibfnamefont {A.}~\bibnamefont
  {Blais}}, \bibinfo {author} {\bibfnamefont {A.~L.}\ \bibnamefont {Grimsmo}},
  \bibinfo {author} {\bibfnamefont {S.~M.}\ \bibnamefont {Girvin}},\ and\
  \bibinfo {author} {\bibfnamefont {A.}~\bibnamefont {Wallraff}},\ }\bibfield
  {title} {\bibinfo {title} {Circuit {{Quantum Electrodynamics}}},\ }\href
  {https://doi.org/10.1103/RevModPhys.93.025005} {\bibfield  {journal}
  {\bibinfo  {journal} {Reviews of Modern Physics}\ }\textbf {\bibinfo {volume}
  {93}},\ \bibinfo {pages} {025005} (\bibinfo {year} {2021})}\BibitemShut
  {NoStop}%
\bibitem [{\citenamefont {Manolatou}\ \emph {et~al.}(1999)\citenamefont
  {Manolatou}, \citenamefont {Khan}, \citenamefont {Fan}, \citenamefont
  {Villeneuve}, \citenamefont {Haus},\ and\ \citenamefont
  {Joannopoulos}}]{manolatou_coupling_1999}%
  \BibitemOpen
  \bibfield  {author} {\bibinfo {author} {\bibfnamefont {C.}~\bibnamefont
  {Manolatou}}, \bibinfo {author} {\bibfnamefont {M.}~\bibnamefont {Khan}},
  \bibinfo {author} {\bibfnamefont {S.}~\bibnamefont {Fan}}, \bibinfo {author}
  {\bibfnamefont {P.}~\bibnamefont {Villeneuve}}, \bibinfo {author}
  {\bibfnamefont {H.}~\bibnamefont {Haus}},\ and\ \bibinfo {author}
  {\bibfnamefont {J.}~\bibnamefont {Joannopoulos}},\ }\bibfield  {title}
  {\bibinfo {title} {Coupling of modes analysis of resonant channel add-drop
  filters},\ }\href {https://doi.org/10.1109/3.784592} {\bibfield  {journal}
  {\bibinfo  {journal} {IEEE Journal of Quantum Electronics}\ }\textbf
  {\bibinfo {volume} {35}},\ \bibinfo {pages} {1322} (\bibinfo {year}
  {1999})}\BibitemShut {NoStop}%
\end{thebibliography}%
	
\end{document}